\documentclass[oldversion]{aa}

\usepackage{txfonts}

\usepackage{graphicx}
\usepackage{longtable}
\begin{document} 

\title{The Gaia-ESO Survey: Reevaluation of the parameters of the open cluster
Trumpler~20 using photometry and spectroscopy\thanks{Based on the data obtained at ESO 
telescopes under programme
188.B-3002 (the public Gaia-ESO spectroscopic survey, PIs Gilmore \& Randich) and on the archive 
data of the programme 083.D-0671.}\fnmsep\thanks{
    Table 3 and the photometric catalogue with differential reddening corrections are only available in electronic form at the CDS via anonymous ftp to {\tt cdsarc.u-strasbg.fr} (130.79.128.5) or via
    {\tt http://cdsweb.u-strasbg.fr/cgi-bin/qcat?J/A+A/???/???}}}
\author{
P. Donati \inst{1,2}
\and
T. Cantat Gaudin\inst{3,4} 
\and
A. Bragaglia\inst{2}
\and
E. Friel\inst{5}
\and
L. Magrini\inst{6}
\and
R. Smiljanic\inst{7,8}
\and
A. Vallenari\inst{4}
\and
M. Tosi\inst{2}
\and
R. Sordo\inst{4}
\and
G. Tautvai\v{s}ien\.{e} \inst{9}
\and
S. Blanco-Cuaresma\inst{10,11}
\and
M.T. Costado\inst{12}
\and
D. Geisler \inst{13}
\and
A. Klutsch \inst{14}
\and
N. Mowlavi \inst{15}
\and
C. Mu\~{n}oz \inst{13}
\and
I. San Roman\inst{13}
\and
S. Zaggia \inst{4}
\and
G. Gilmore\inst{16}
\and
S. Randich\inst{6} 
\and
T. Bensby\inst{17}
\and
E. Flaccomio\inst{18}
\and
S.~E. Koposov\inst{16,19}
\and
A.~J. Korn \inst{20}
\and
E. Pancino \inst{2,21}
\and
A. Recio-Blanco \inst{22}
\and
E. Franciosini \inst{6}
\and
P. de Laverny \inst{22}
\and
J. Lewis\inst{16}
\and
L. Morbidelli\inst{6}
\and
L. Prisinzano \inst{18}
\and
G. Sacco\inst{6}
\and
C.~C. Worley\inst{16}
\and
A. Hourihane\inst{16}
\and
P. Jofr\'e \inst{16}
\and
C. Lardo\inst{2}
\and
E. Maiorca\inst{6}
}
\institute{
Dipartimento di Fisica e Astronomia, Universit\`a di Bologna, via Ranzani 1, 40127 Bologna, Italy
\and
INAF-Osservatorio Astronomico di Bologna, via Ranzani 1, 40127 Bologna, Italy
\and
Dipartimento di Fisica e Astronomia, Universit\`a di Padova, vicolo Osservatorio 3, 35122 Padova, Italy
\and
INAF-Osservatorio Astronomico di Padova, vicolo Osservatorio 5, 35122 Padova, Italy 
\and
Department of Astronomy, Indiana University, Bloomington, IN 47405, USA 
\and
INAF-Osservatorio Astrofisico di Arcetri, Largo E. Fermi, 5, 50125 Firenze, Italy 
\and
European Southern Observatory, Karl-Schwarzschild-Str. 2, 85748 Garching bei M\"{u}nchen, Germany 
\and
Department for Astrophysics, Nicolaus Copernicus Astronomical Center, ul. Rabia\'{n}ska 8, 87-100 Toru\'n, Poland
\and
Institute of Theoretical Physics and Astronomy, Vilnius University, Go\v{s}tauto 12, LT-01108 Vilnius, Lithuania 
\and
Univ. Bordeaux, LAB, UMR 5804, F-33270, Floirac, France
\and
CNRS, LAB, UMR 5804, F-33270, Floirac, France
\and
Instituto de Astrof\'isica de Andaluc\'ia, CSIC, Apdo 3004, 18080 Granada, Spain
\and
Departamento de Astronom\'{i}a, Universidad de Concepci\'{o}n, Casilla 160-C, Concepci\'{o}n, Chile
\and
INAF - Osservatorio Astrofisico di Catania, via S. Sofia 78, 95123 Catania, Italy
\and
Astronomy Department, University of Geneva, Switzerland
\and
Institute of Astronomy, University of Cambridge, Madingley Road, Cambridge CB3 0HA, UK
\and
Lund Observatory, Department of Astronomy and Theoretical Physics, Box 43, SE-221\,00 Lund, Sweden
\and
INAF - Osservatorio Astronomico di Palermo, Piazza del Parlamento 1, 90134, Palermo, Italy
\and
Moscow M.V. Lomonosov State University, Sternberg Astronomical Institute, Universitetskij pr., 13, 119992 Moscow, Russia
\and
Department of Physics and Astronomy, Uppsala University, Box 516, SE-75120 Uppsala, Sweden
\and
ASI Science Data Center, Via del Politecnico SNC, 00133 Roma, Italy
\and
Laboratoire Lagrange (UMR7293), Universit\'{e} de Nice Sophia Antipolis, CNRS, Observatoire de la C\^{o}te d'Azur, BP 4229, F-06304 Nice cedex 4, France 
}

\titlerunning{Trumpler 20}
\authorrunning{Donati et al.}

   \date{Accepted on 9 Dec 2013}


 \abstract
    {Trumpler~20 is an old open cluster (OC) located toward the Galactic centre, at 
about 3 kpc from the Sun and $\sim$7 kpc from the Galactic centre. Its position makes this cluster particularly interesting in the framework of 
the chemical properties of the Galactic disc because very few old OCs reside in the inner part of the 
disc. For this reason it has been selected as a cluster target of the Gaia-ESO Survey, and spectra of 
many stars in the main sequence and red clump phases are now available. Moreover, although it has 
been studied by several authors in the past, no consensus on the evolutionary status of Tr~20 has 
been reached. The heavy contamination of field stars (the line of sight of Tr~20 crosses the Carina 
spiral arm) complicates a correct interpretation. Another interesting aspect of the cluster is that it 
shows a broadened main-sequence turn-off and a prominent and extended red-clump, characteristics 
that are not easily explained by classical evolutionary models. Exploiting both spectroscopic 
information from the Gaia-ESO Survey (and the ESO archive) and literature photometry, we obtain a 
detailed and accurate analysis of the properties of the cluster. We make use of the first accurate 
metallicity measurement ever obtained from several spectra of red clump stars, and of cluster 
membership determination using radial velocities. According to the evolutionary models adopted, we find that Tr~20 has an age in the range 1.35-1.66 Gyr, an 
average reddening $E(B-V)$ in the range 0.31-0.35 mag, and a distance modulus $(m-M)_0$ between 12.64 and 12.72 mag. The spectroscopic metallicity is [Fe/H]=+0.17 dex. We discuss the structural properties of the object and constrain 
possible hypotheses for its broadened upper main sequence by estimating the effect of differential 
reddening and its extended red clump.   
}


   \maketitle


\section[]{Introduction}\label{sec:intro}
The Gaia-ESO Survey \citep[GES, see][]
{gilmore,rand13} is a large, public spectroscopic survey of the Galaxy using the high-resolution multi-object spectrograph 
FLAMES (see \citealt{pasquini02}) on the Very Large Telescope (ESO, Chile). It targets about 10$^5$ stars and covers 
the bulge, thick and thin discs, and halo components, as well as a sample of about 100 open clusters (OCs) 
of all ages, metallicities, locations, and masses.
While the Gaia-ESO Survey will leave an unprecedented legacy for high-resolution spectroscopic observations, its value can even be increased by synergies with other missions. In the long run, the Gaia satellite will produce distances 
and proper motions for all objects targeted during the Gaia-ESO Survey, which has in fact been conceived also as 
a ground-based complement to Gaia.
On shorter timescales, information present in the archives and literature plays a fundamental role in 
enhancing both the efficiency of the spectroscopic observations and the scientific return of the survey. 
For instance, photometry of the Gaia-ESO Survey targets is essential for a full understanding of their physical 
parameters. In this framework we present a comprehensive and homogeneous analysis of the 
archive photometry and the Gaia-ESO Survey spectroscopy of one of the Gaia-ESO Survey OC targets: Trumpler~20. 

The open cluster Trumpler~20 (Tr~20) is a relatively old OC located in the fourth quadrant of the 
Galactic plane (RA=12:39:32, Dec=-60:37:36, \citealt{seleznev10}; l=301.475$^{\circ}$, b=2.221$^
{\circ}$, \citealt{dias02}). Only a few studies are available (see Sect. \ref{data}). 
There is no consensus on the evolutionary status of Tr~20: its position in the Galactic disc is 
such that many field interlopers pollute the main evolutionary phases on the colour-magnitude 
diagram (CMD), jeopardising the derivation of the cluster parameters. Moreover, the reddening, and 
in particular the differential reddening (DR) across the face of the cluster, plays a considerable role in shaping the CMD morphology \citep[see e.g.,][]{pla08,pla12}. Tr~20 shows a 
peculiar morphology of the red-clump (RC) phase, common to other OCs 
\citep[see e.g.,][]{merm89,merm98,gir2000b}, but still little explained and hardly understood (see Sect. \ref
{sec:rc} for a discussion). Finally, prior to the Gaia-ESO Survey data, there has been no systematic study of 
the metallicity using spectroscopic analysis (the value of [Fe/H]$=-0.11$ dex by \citealt{pla08} is 
based on a single star). 

In the context of the Gaia-ESO Survey, Tr~20 is the first old OC observed: its proximity (about 3 kpc from the Sun, 
according to \citealt{pla08} and \citealt{carraro10}), its mass, its age, and its position (located inside the solar ring in 
the direction of the Galactic centre) make this cluster very interesting.
In fact only very few OCs older than 1 Gyr are 
known \citep[less than 20\%, see e.g.,][and web updates]{dias02}, but they are 
ideal probes for Galactic disc chemical evolution and structure studies \citep[see e.g.,][]
{friel95,bt06,magrini09,pancino10,lepine11,yong12}. Furthermore, few OCs in the inner disc are 
known. It is fundamental that not only the metallicity and detailed chemistry of Tr~20 are 
accurately measured, but also its age and distance.

The selection of targets observed by the Gaia-ESO Survey is different for UVES and GIRAFFE instruments: whilst UVES fibres are preferentially allocated to high-probability cluster members, the selection of Giraffe targets, mainly based on photometric criteria, is unbiased and high priority is given to all candidate members (see Bragaglia et al., in preparation, for more details). One of the 
goals of the Gaia-ESO Survey is to determine the membership using the radial velocity (RV) and, if possible, 
exploiting the additional information on the chemical abundances. 

\begin{figure}
\begin{center} \includegraphics[scale=0.45]{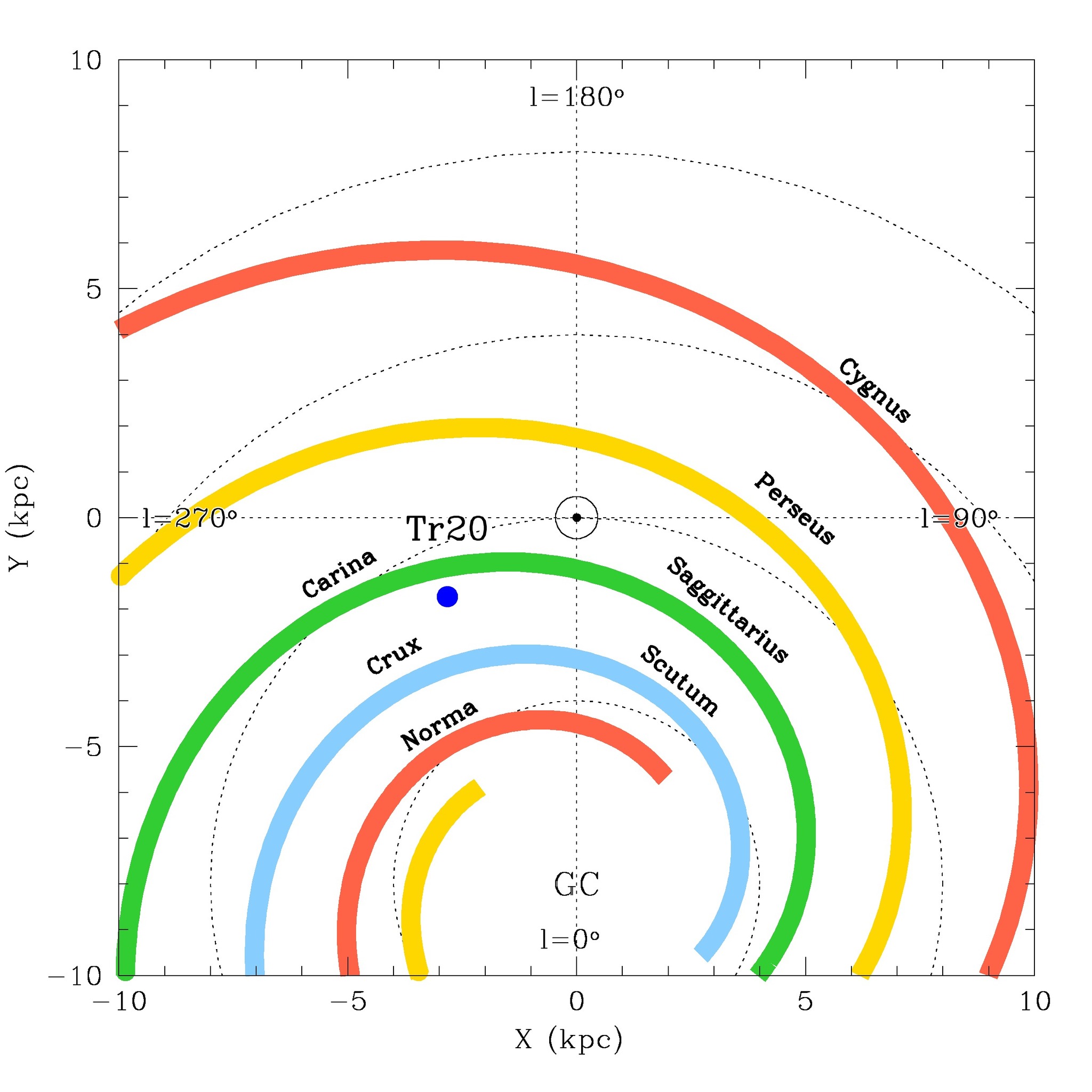} \caption{\label{fig:mw1}View of the MW based 
on the model of Vall{\'e}e et al. (2005). 
The position of Tr~20, indicated with a filled blue circle, is based on the results of this analysis. 
The $\odot$  symbol at the centre of the figure marks the position of the Sun, while the different approximate position of the spiral arms are marked with coloured stripes.} 
\end{center}
\end{figure}

The goal of this paper is to determine with better accuracy the cluster parameters from
 the best-fitting isochrones of the latest evolutionary models,  taking into 
account the effect of differential reddening and the spectroscopic information 
from the Gaia-ESO Survey for cluster membership and chemical abundance. 
The GIRAFFE spectra are exploited  for determining the RV distribution of the clusters targets, 
the UVES spectra mainly for the good constraint to the metallicity of the cluster provided by
accurate chemical abundances analysis. The large and homogeneous dataset of the 
survey guarantees a first comprehensive 
study of several OCs using the same methods. Among intermediate-age and old OCs, Tr~20, NGC~4815 and NGC~6705 (M~11) are also currently being
analysed: Magrini et al. (to be submitted) discuss their 
chemical abundances in general, while NGC~6705 is studied in Cantat-Gaudin et al. (in prep.), and NGC~4815  in Friel et al. (in prep.). 

This paper is organised as follows: in the second section we summarise previous studies 
on Tr~20. In Section 3 we briefly describe the photometric and spectroscopic data sets. Sections 4 
and 5 describe the 
analysis of the photometric and spectroscopic data, focusing on the estimate of the differential 
reddening, the peculiar morphology of the RC, the cluster metallicity and radial velocity distribution. 
The cluster parameters are described in Section 6, while a summary of the whole analysis is 
presented in Section 7.

\section[]{Trumpler 20: previous studies}\label{data}
The first study of Tr~20 that defined its evolutionary status using photometry and isochrones was made 
by \citet[][hereafter MS05]{mcswain05}. Their goal was to determine the fraction of Be stars relative to 
B stars as a function of cluster age. They used Str\"omgren photometry and fitted isochrones to identify 
the B stars in each cluster. For Tr~20 they found only one possible Be star and very few B stars. 
However, they apparently missed the cluster signatures and fitted a too young isochrone (160 Myr) to the brighter stars, which are more likely field stars.

\citet[][hereafter P08]{pla08} used $BVI$ photometry on a 20$\arcmin\times$20$\arcmin$ field of 
view, reaching a magnitude of $V=18.5$. The main sequence (MS) of the cluster is clearly visible together 
with the RC. They estimated an age of 1.3 Gyr using the Padova isochrones \citep{gir2000} for 
$Z=0.015$ (the solar composition for this isochrone set is $Z=0.019$), a distance modulus 
of $(m-M)_0=12.6$ mag, and a reddening $E(B-V)=0.46$ mag. They observed three 
red giant branch (RGB) 
and three RC stars with FEROS ($R=48000$); five were found to be members based on RV. One of 
the spectra had a sufficient signal-to-noise ratio (S/N) to derive stellar parameters and abundances, 
from which they obtained a metallicity of [Fe/H]=-0.11.

\citet[][hereafter S10]{seleznev10} derived the cluster parameters and structural 
parameters using $V$ and $I$ photometry on a field of view of 13.5$\arcmin\times$13.5$\arcmin$, 
reaching $V\sim22$ mag. They estimated an age of 1.5 Gyr, $E(B-V)=0.48$ mag, and
 $(m-M)_0=12.4$ mag ($R_{GC}=7.3$ kpc) using the Padova isochrones of solar metallicity 
 \citep{gir2000}. They 
provided a measurment of the radius and centre of the cluster using both their optical photometry and 
infrared photometry (from the 2MASS catalogue, \citealt{2mass}, for a larger field of view). They derived a 
radius of $r=5.4\arcmin$ and centre coordinates RA=12:39:32 and Dec=-60:37:36.  

\citet[][hereafter C10]{carraro10} observed Tr~20 using $UBVI$ filters, deriving the cluster 
parameters using both two-colour diagrams and CMD. They found an age of 1.4 Gyr from the solar 
metallicity Padova models \citep{gir2000}. This result is consistent with the one found by P08, but for 
a different reddening value: the distance modulus C10 measured is $(m-M)_0=12.6$ mag and the 
reddening is $E(B-V)=0.35$ mag. They explained the visible broadening of the main sequence turn-off 
(MSTO or simply TO) with the strong impact from binary systems and the unavoidable contamination by field 
interlopers. They also discussed the prominent and extended RC of the cluster. 

\citet[][hereafter P12]{pla12} estimated the radial velocities of nearly 1000 stars belonging to the 
upper MS and RC/RGB using the GIRAFFE fibres on the FLAMES instrument at VLT. Neither 
photometry nor RVs are public. They derived an average  $<RV>=-40.40\pm0.12$ km~s$^{-1}$ using 
68 RC stars. They were able to define 471 cluster members and suggested that about 50 to 100 stars might still 
be field stars when the statistics on the rotation velocity $v\sin i$ is also taken into account. They also 
estimated the differential reddening, concluding that it plays a strong role in shaping the cluster CMD 
morphology. In particular, they ruled out the hypothesis of multiple populations in Tr~20 (which they 
proposed in earlier works) as an explanation of the broad MSTO.

We summarise the results available in the literature in Table \ref{tab:lit}, where we list age, metallicity,
the method used 
to derive metallicity (from spectroscopy, S, or photometry, P), the distance modulus and reddening, the distance from the Sun and the 
Galactic centre ($R_{GC}$ has been computed assuming $R_{GC,\odot}=8$ kpc, 
see \citealt{mal13}), and the height above the Galactic plane. 
Regarding the age, a good consensus is 
obtained within the quoted errors, except for MS05, for the reason explained above. The differences 
in distance modulus and reddening are related to the age and metallicity adopted and to differences 
in the photometric data. In the light of the new results obtained within the Gaia-ESO Survey survey, the metallicity 
obtained with high-resolution spectroscopy is of great importance also to determine the other parameters more accurately and limit the degeneracy between them. We used this 
information and took into account the differences in photometry and different evolutionary models to 
derive a reliable estimate of age, distance, and reddening of Tr~20. We also improved the available 
photometric catalogue, proposing a correction for differential reddening that helps to explain the 
morphology of the MSTO and better constrain the evolutionary status of the cluster. Moreover, we discuss in 
detail the peculiar RC, which is too extended with respect to the predictions of standard stellar evolutionary models.

\begin{table*}
\centering
\caption{ \label{tab:lit} Summary of the parameters estimated for Tr~20 by different authors.}
\begin{tabular}{lccccccccr}
\hline
\hline
Refs & age & [Fe/H] & method & $(m-M)_0$ & $E(B-V)$ &  $d_{\odot}$ & $R_{GC}$ & Z \\
       & (Gyr) & (dex) & & (mag) & (mag) & (kpc) & (kpc) & (pc) \\ 
\hline
 MS05 & 0.16 & - & - &11.92 & 0.26 & 2.42 & 7.05 & 93.82  \\
 P08 & 1.30 & -0.11 & S & 12.60 & 0.46 & 3.31 & 6.88 & 128.33 \\
 S10  & 1.50 & 0.0 & P & 12.40 & 0.48 & 3.02 & 6.92 & 117.03 \\
 C10  & 1.40 & 0.0 & P & 12.60 & 0.35 & 3.31 & 6.88 & 128.33  \\
\hline
\end{tabular}
\end{table*}

\section{Observational data}
\subsection{Photometry}\label{sec:lit}
\begin{figure}
\begin{center} \includegraphics[scale=0.45]{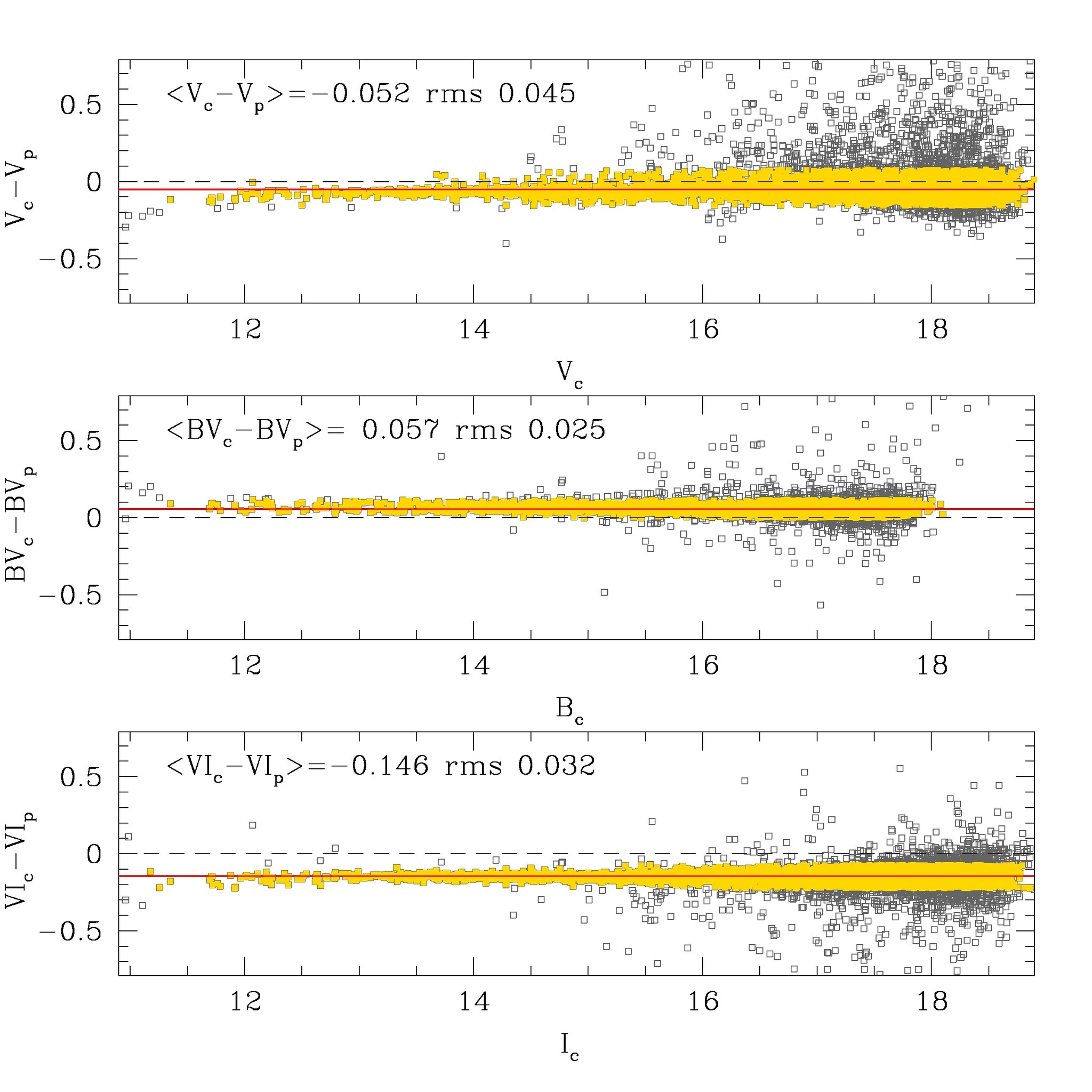} \caption{\label
{fig:compareV}Systematic differences between the data sets of P08 and C10 in $V$, $(B-V)$, and
$(V-I)$. The yellow filled squares were used to compute the average differences.} 
\end{center}
\end{figure}

We obtained the $BVI$ photometry of P08 for Tr~20 from the CDS\footnote{The Strasbourg 
Astronomical Data Center, see http://cds.u-strasbg.fr/.} and that of C10 from WEBDA\footnote{The 
on-line database collecting OC photometry, see webda.physics.muni.cz.}.
By using the {\sc catapack}\footnote{Made available by Paolo Montegriffo at the INAF Bologna 
Observatory.} programme, we were able to cross-identify the stars in common between the two catalogues 
and compare of their photometry. 
The difference between P08 and C10 is on average $-0.052 \pm 0.045$ in $V$, $+0.057 \pm 0.025$ 
in $(B-V)$, and $-0.146 \pm 0.032$ in $(V-I)$. Fig. \ref{fig:compareV} shows these 
offsets and how they scale with magnitude.

Note that our result is different from that in C10 (their Fig. 3); the $V$ magnitudes 
and $(B-V)$ colours from the two works roughly agree, but we were unable to reproduce 
the perfect 
agreement for $(V-I)$. We have no apparent explanation for this and decided to give preference 
to the $(B-V)$ colour with respect to the $(V-I)$ one throughout this analysis. Moreover, these 
differences have an important impact on the derivation of the cluster properties, as discussed in 
Sect. \ref{sec:param}.

\begin{figure*}
\begin{center} 
\includegraphics[scale=0.8]{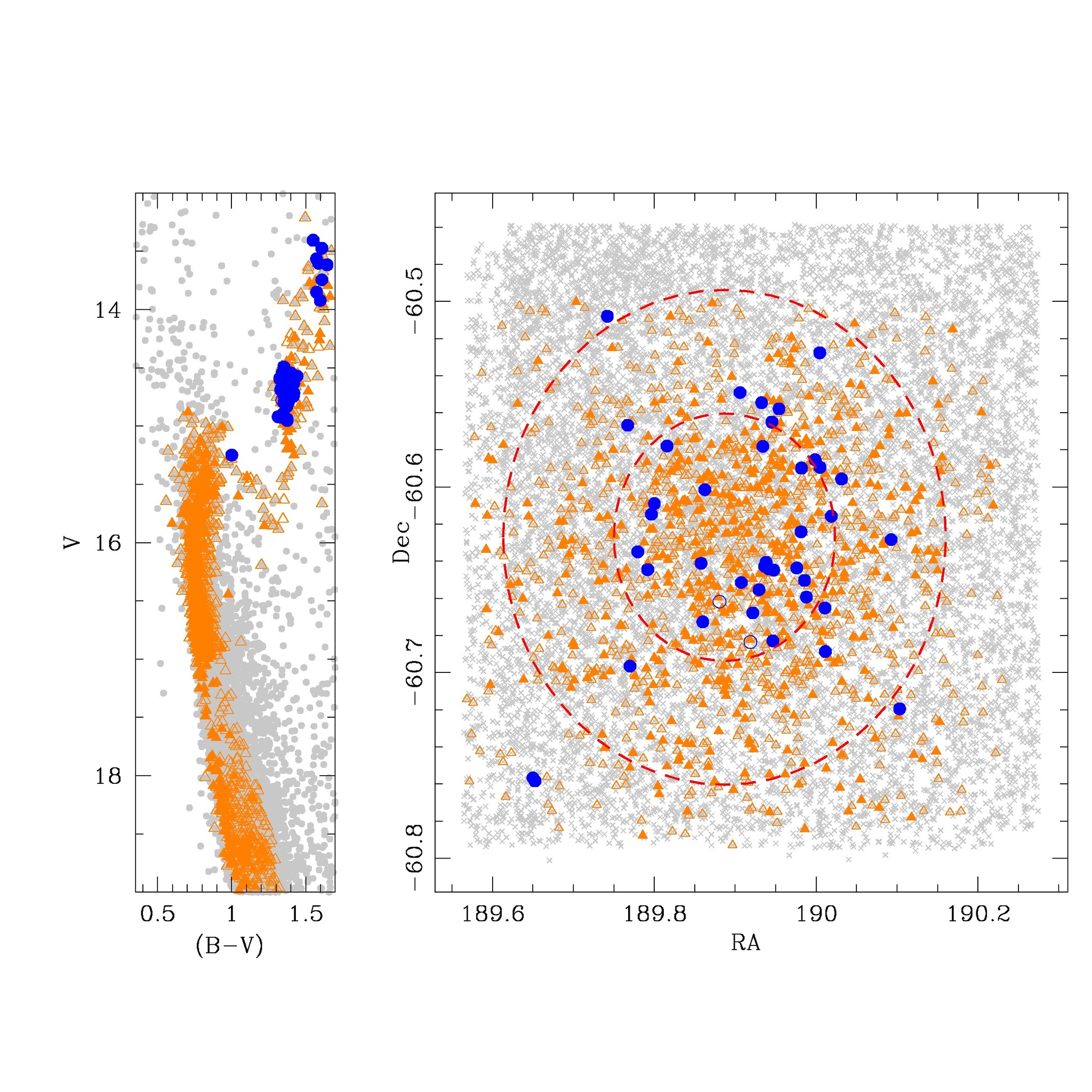} \caption{\label{fig:gestgts}\textit
{Left panel:} CMD of the GES targets for Tr 20 with a distance from the  centre $d<8\arcmin$. 
\textit{Right panel:} Spatial 
distribution of the GES targets; the red dashed circles define the regions for $d<4\arcmin$ and 
$d<8\arcmin$ from the centre. In both panels, the orange triangles are GIRAFFE targets, while the 
blue filled circles are those of UVES; filled symbols are candidate members for 
RV (see Sect. 5). The stars in the photometric catalogue are plotted in grey.}
\end{center}
\end{figure*}

\subsection{Spectroscopy}\label{sec:spectro1}
Many stars of Tr~20 have been observed within the Gaia-ESO Survey, using the GIRAFFE HR15n grating and the 
UVES 580nm setup. The targets were selected to lie on or near the evolutionary sequences of the CMD. Concerning the cluster MS, stars with colour between the blue envelope of the MS and about 0.3 mag redder were considered potential targets, with a higher priority given to stars closer to the cluster centre. The cluster RC is evident and rich, and all stars in this phase were considered potential targets. Some potential sub giant stars were also taken into account. The GIRAFFE targets were chosen 
mainly on the MS and on the probable subgiant branch (SGB), the RGB, and the RC, while the UVES 
targets fall on the RC. In Fig. \ref{fig:gestgts} we show their spatial position and their locus on the 
CMD. The S/N ratio of the spectra depends mainly on the luminosity of the targets, 
spanning from  about 10 for the fainter targets up to 300 for the most luminous. 
The median S/N is about 30 for the GIRAFFE targets, and 60 for those of UVES.

The Gaia-ESO Survey observations of Tr~20 have been obtained in Spring 2012 and 2013. We observed 42 stars on the RC with the UVES fibres, while 527 MS and giant stars were observed 
with the GIRAFFE fibres. Only 13 UVES targets were fully analysed as part of the first 
internal data release (GESviDR1Final) of the survey, which encompassed the first six months of 
observations, while for the other targets a full analysis will be available in the next releases. However, 
the RV is available for all the Gaia-ESO Survey stars and for the 954 archive spectra taken by Platais and 
collaborators using the GIRAFFE HR09b grating, as described in P12 and re-analysed inside the Gaia-ESO Survey; this greatly benefits the analysis of the RV distribution described in this paper.
We were able to cross-identify 953 of them. In our sample, 40 stars were observed both 
with UVES and GIRAFFE; 110 
stars with the two GIRAFFE setups; 10 with all the setups. More information on the observations is 
given in Table \ref{tab:sumobs}. In Table 3, only available online, we list the relevant information for all the spectra (obtained with the three different setups HR09b, HR15n, and UVES) of the targets used in this paper. We report the identification number from the C10 and P08 catalogues, the Gaia-ESO Survey id, RA, and Dec coordinates, the magnitudes $V$ and $B$ from C10 when available, otherwise those from P08, and the RVs.

The Gaia-ESO Survey consortium is structured into several working groups, WGs, with specific duties. The data 
reduction is performed by WG 7 and a comprehensive description of the methods used can be found 
in Sacco et al. and Gilmore et al. (both in prep.). 
The RV information,  available for all the targets observed with both GIRAFFE and UVES, is determined by WG 8 (see Gilmore et al., in prep.). 
For the abundance derivation, see
Sec.~\ref{sec:spectra};
at the moment, only 13 UVES targets are fully analysed and were used here, while those of GIRAFFE will be released in the near future.
With this new spectroscopic information it is possible to derive the cluster parameters with unprecedented accuracy.

\begin{table}
\centering
\caption{ \label{tab:sumobs} Summary of the GES and public spectroscopic observations for Tr~20.}
\begin{tabular}{l l l r}
\hline
\hline
Setup used   & $\lambda\lambda$ & Time exp. & \# stars \\
                     &(nm)                        &(min)         & \\
\hline
GIRAFFE HR15n  & 476.0-684.0         & 3$\times$50  & 525  \\ 
UVES 580nm        & 647.0-679.0   & 3$\times$50  & 42 \\ 
GIRAFFE HR09b  & 514.3-535.6   & $^a$             & 954 \\
\hline
\end{tabular}

$^a$ESO public archive data of programme 083.D-0671 (see P12) processed within the GES 
\end{table}

\section[]{Photometric analysis}\label{sec:photo}
\subsection[]{Centre, mass, and radius}\label{sec:king}
We made a broad photometric selection to remove very obvious field polluters from the sample, keeping the 
MS, MSTO, and RC stars. We determined the position of the cluster centre by an 
iterative process using the same method as described in \cite{donati12}. We computed the barycentre of the positions of the stars, then took the 70\% of 
stars closest to this position and recomputed their barycentre, and iterated until convergence on a 
central position.
To avoid selecting too many field stars we set magnitude cuts at $V=16$, 17, and 18. This led to 
similar results because the position we obtain is identical within $0.5\arcmin$ around the coordinates: 
$\rm{RA}=12^{\rm{h}} 39^{\rm{m}} 32\fs8$, $\rm{DEC}=-60\degr 37\arcmin 37\farcs 4$ (or, in 
Galactic coordinates: $l=301\degr.47$, $b=2\degr.21$). 

\begin{figure}
\begin{center} \includegraphics[scale=0.45]{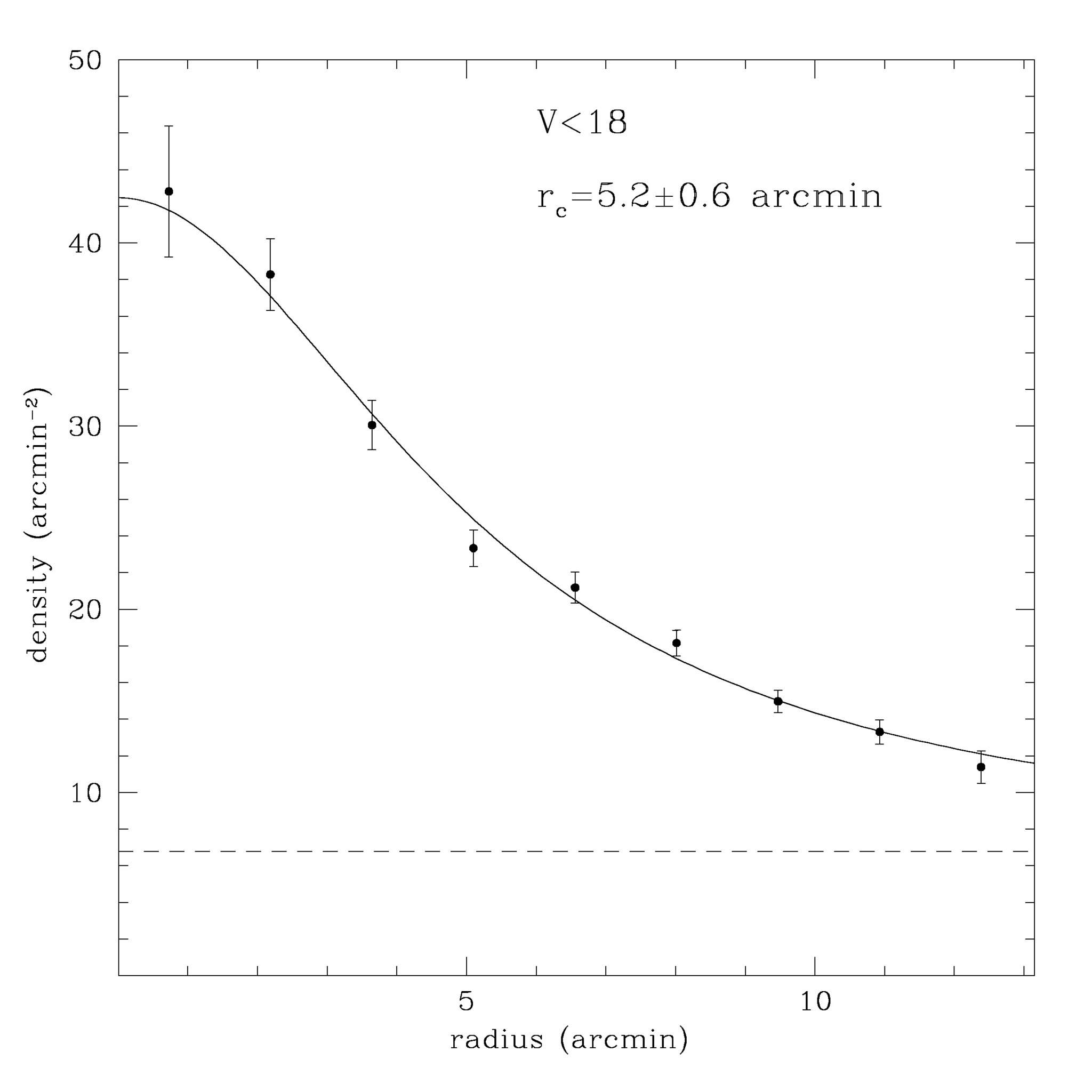} \caption{\label{fig:kingprof}Density 
profile of Tr~20 using stars with $V<18$. The error bars are the random errors. The best-fit is a two-parametric King model of core radius $5.2 \pm 0.6$ arcmin.} \end{center}
\end{figure}

Tr~20 is densely populated and stands out against the field stars, which enabled us to follow its 
density profile. C10 indicated that the completeness of their photometry is better than 90\% for 
magnitudes $V<19$. To be conservative, we only used 
stars brighter than $V=18$. We performed a least-squares fit of a two-parameter King profile \citep
{king62},
\begin{equation} \label{eq:kingProf}
f(r) = \rho_{bg} + \frac{ \rho_{0} }{ 1+(r/r_{c})^{2} },
\end{equation}
where $\rho_{bg}$ the background density, $\rho_{0}$ the central density, and $r_{c}$ the core 
radius are left as free parameters. The observed profile and the best-fit are shown in Fig. \ref
{fig:kingprof}. Using a three-parameter King model that also takes into account a tidal radius does not improve the goodness of fit. This means that the tidal radius of Tr~20 is larger than our field of view, and the region where the density profile starts to decrease faster is too far out for our data.

Since the model fitting provides a value for the background stellar density, we were able to remove its 
contribution to the star counts. The density profile was integrated to obtain the total number of stars 
contained in the cluster. Assuming a Salpeter mass function and the best-fitting PARSEC isochrone 
(see Sect.~\ref{sec:param}), we produced a synthetic population that contains the same number of 
stars as Tr~20 in the magnitude range $V<18$. We added all the masses of stars down to $0.08$ 
$M_{\odot}$ and found a total mass of about $6\,800$ $M_{\odot}$.

Varying the age of the isochrone within the uncertainties gives an estimate of the error on the total 
mass. This whole operation was also performed by selecting stars brighter than $V=17$ and $V=16$, which yelded 
very similar results. Finally, we estimated the total mass of Tr~20 to be $6\,700 \pm 800$ $M_{\odot}$.
We can draw a more conservative estimate by adding the contribution of the random errors to the 
star counts. Considering the Poissonian uncertainty, we find masses between the extreme values of 
5500 and 8000 $M_{\odot}$.

\subsection[]{Differential reddening}\label{sec:dr}
The position of Tr~20 in the Galactic disc and its high reddening estimates strongly suggest 
differential reddening (DR, as was discussed, e.g., by P12). The main effect of DR on the 
CMD appearance is that it broadens the sequences. This is mainly due to the presence of patchy dust 
structures in the field of view, which cause different extinctions along the line of sight. Photometric errors 
have a similar effect on the CMD appearance, but the broadened MS of Tr~20 cannot be explained 
only with errors since they are too small in the two photometric studies considered, as discussed in 
the original papers.

Other explanations cannot be a priori ruled out. For example, a significant age spread during the star 
formation process can produce an observed broadening in the TO phase and in later phases, and 
unresolved binary systems also widen the MS and TO since they have redder colour and brighter 
luminosity than single objects. A metallicity spread has in principle an effect very similar to 
the DR, even though there is no evidence among the OCs of inhomogeneities in the overall metallicity. 
Finally, stellar rotation might also affect colour and magnitude; this would make rotating stars seem fainter and redder 
\citep[see][]{bast09,li12}. Hot early-type stars, such as those near the MSTO, can be fast rotators, as 
demonstrated by P12 and as seen from the Gaia-ESO Survey GIRAFFE spectra. However, the rotation has a mild 
effect on the CMD appearance (see P12), confirming the study of \cite{gir11}, who excluded rotation as a possible explanation of the extended TOs observed sometimes.

An estimate of the DR for Tr~20 has been performed in P12. They used about 200 slow-rotator stars 
(located in the upper MS) and evaluated their distance, along the reddening vector direction, from a 
hand-defined blue envelope of the MS. They smoothed the measurements on a grid using a scale of 
1$\arcmin$ and adopting for each bin the median of the nearest few measurements that fell in the 
same bin of the grid. They demonstrated that the effect of DR on the cluster face is not negligible, but we were unable to use their individual DR values because they are not publicly available at the moment.

We decided to apply a different method of evaluating the DR here. This method is a revision of the one 
described in \cite{mil12} adapted to the case of the OCs, which are less populated and more polluted 
by field stars than the globular clusters. The main steps of the process are the following: 
\begin{itemize}
\item we define a fiducial line along the MS to be used as a reference locus for the DR estimate (the 
choice made for this cluster is described in the following paragraphs);
\item we define a region around the fiducial on the MS (we call it MS box for conciseness): all stars 
falling in this region are used to estimate the DR;
\item for each star in the catalogue we pick up its 30 nearest and brightest stars inside the MS box 
and compute their median distance from the fiducial line in the CMD plane. This distance is used to 
correct colour and magnitude along the reddening vector direction;
\item after the first provisional estimate of the DR is applied star-by-star, we repeat this procedure 
until convergence is reached. The convergence criterion is a user-defined percentage of stars for 
which the DR correction estimate is lower than the average rms on these estimates; 
\item after a final value for the DR is obtained for each star, a binning is performed in the spatial 
plane. The spatial scale used is compatible with the average distance of the 30 neighbour stars 
selected and used for the DR estimate. At this point a rejection of outliers is performed: stars for 
which the DR estimate has an rms higher than average or for which the average distance of the 30 
neighbours is larger than average are not taken into account;
\item a final and reliable value for the DR is then computed as the average value of the DR corrections 
associated with the stars falling in the same bin, and the error on this estimate is the associated rms. 
The DR values obtained in this way are not absolute values, but are relative to the fiducial line.  
\end{itemize}

\begin{figure}
\begin{center} \includegraphics[scale=0.45]{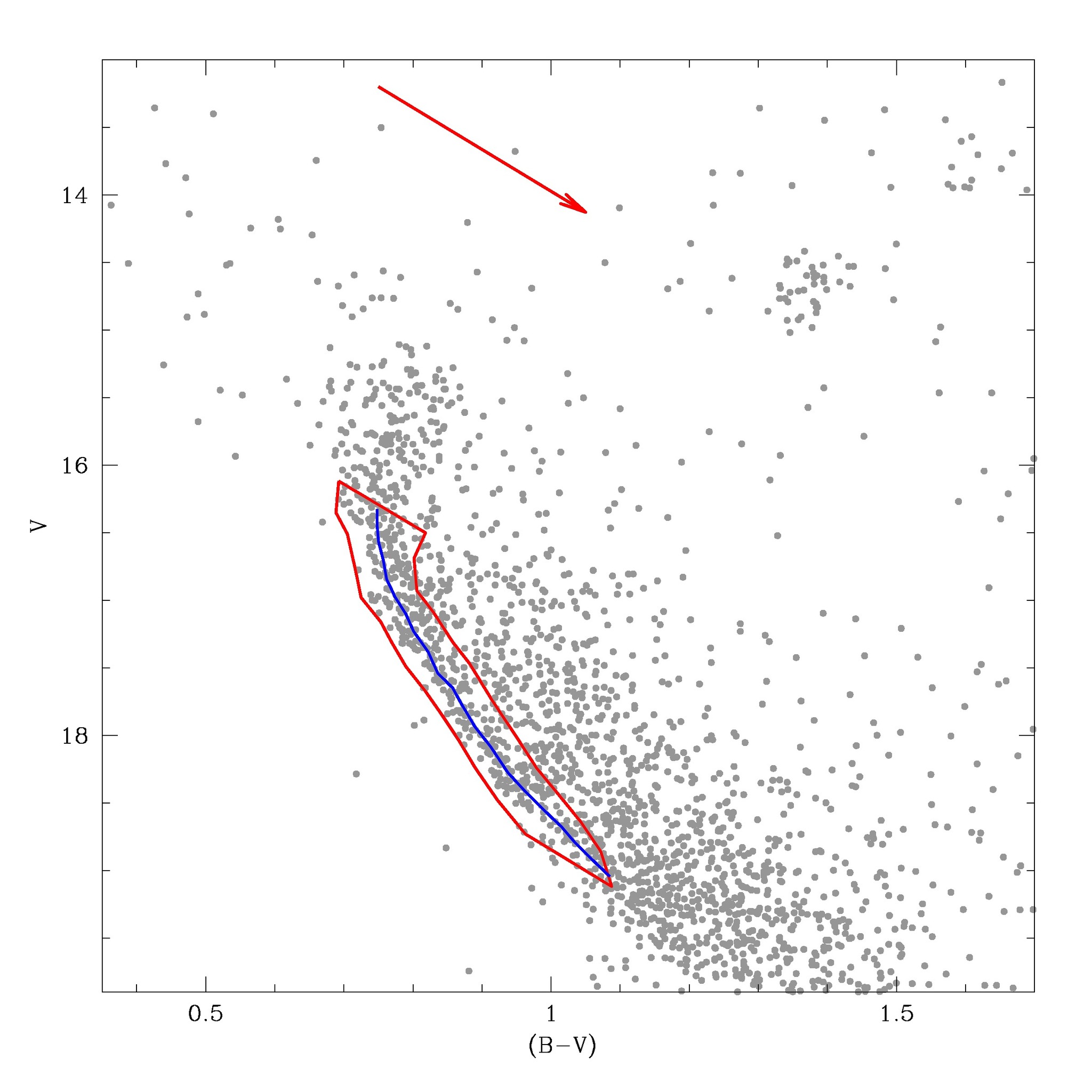} \caption{\label{fig:cmd_box}CMD of Tr~20 
inside $4\arcmin$. The red box and the blue line indicate the MS box and the fiducial line for the DR 
estimate. The red arrow indicates the reddening vector, both in direction and size.} \end{center}
\end{figure}

We used the photometry of C10 because it reaches fainter magnitudes, so that the MS is well described on 
a wider magnitude range. We estimated the DR in the $B-V$ colour, since the C10 and P08 
photometric data agree better in this colour than in $V-I$ (see Sect. \ref{sec:lit}). The 
direction of the reddening vector was derived assuming the standard extinction law ($R_V=3.1$) described in \cite{dean78}. The fiducial line was defined using the CMD of the 
inner part of the cluster (all the stars in C10 inside 4$\arcmin$ to clearly identify the cluster 
signature from the field contamination) and was chosen as the ridge line along the MS. In Fig. \ref{fig:cmd_box} the box and the fiducial line used are highlighted. Several attempts were made to avoid fiducial lines that during estimating the DR led to corrections 
that artificially and noticeably changed the magnitude and colour of the age-sensitive indicators. We
aimed to keep the RC, MSTO, and the blue envelope of the MS as close as 
possible to the observational CMD to limit spurious interpretations of the cluster parameters 
due to DR corrections. When defining the MS box, we avoided the broadened and curved region of 
the TO, where the morphology might hamper a correct interpretation, and the fainter part of the 
MS, where the photometric error is larger.

Taking into account the star counts of the inner and outer parts of the cluster (see Sect. \ref
{sec:king}), we decided to limit the application of correction for DR to stars inside a region of $6\arcmin
$ of radius. For the outer parts the contamination of field stars became significant (the contrast 
density counts with respect to the field plateau drops below 50\%) and any attempt to estimate 
the DR was severely affected by field interlopers. 

For all the stars inside $6\arcmin$, a value of DR was computed using the 30 nearest stars falling in 
the  MS box. Then spatial smoothing was applied to obtain a more robust statistic, adopting a binning of 
$50\arcsec$ in right ascension and declination.

\begin{figure}
\begin{center} 
\includegraphics[scale=0.5]{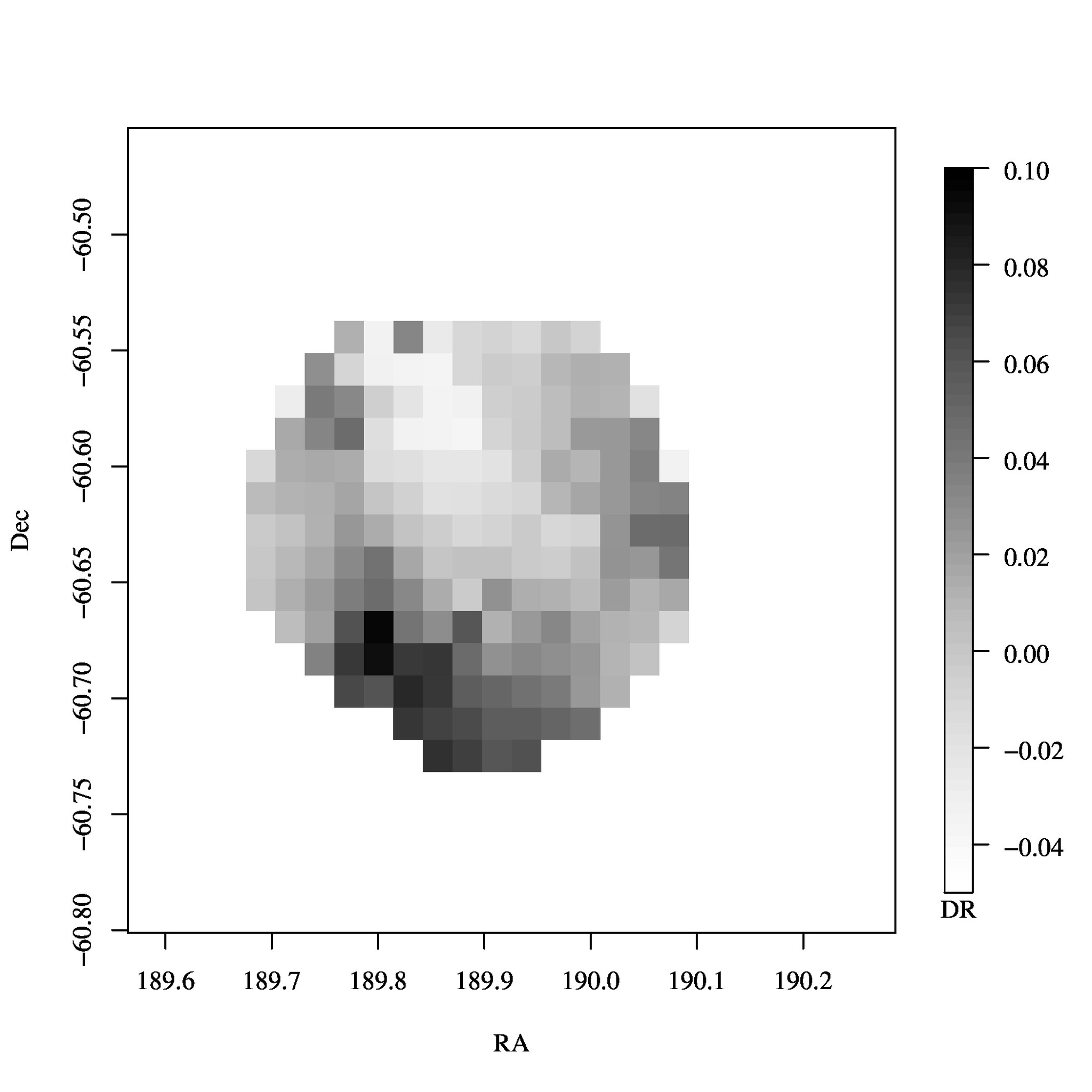}
\includegraphics[scale=0.5]{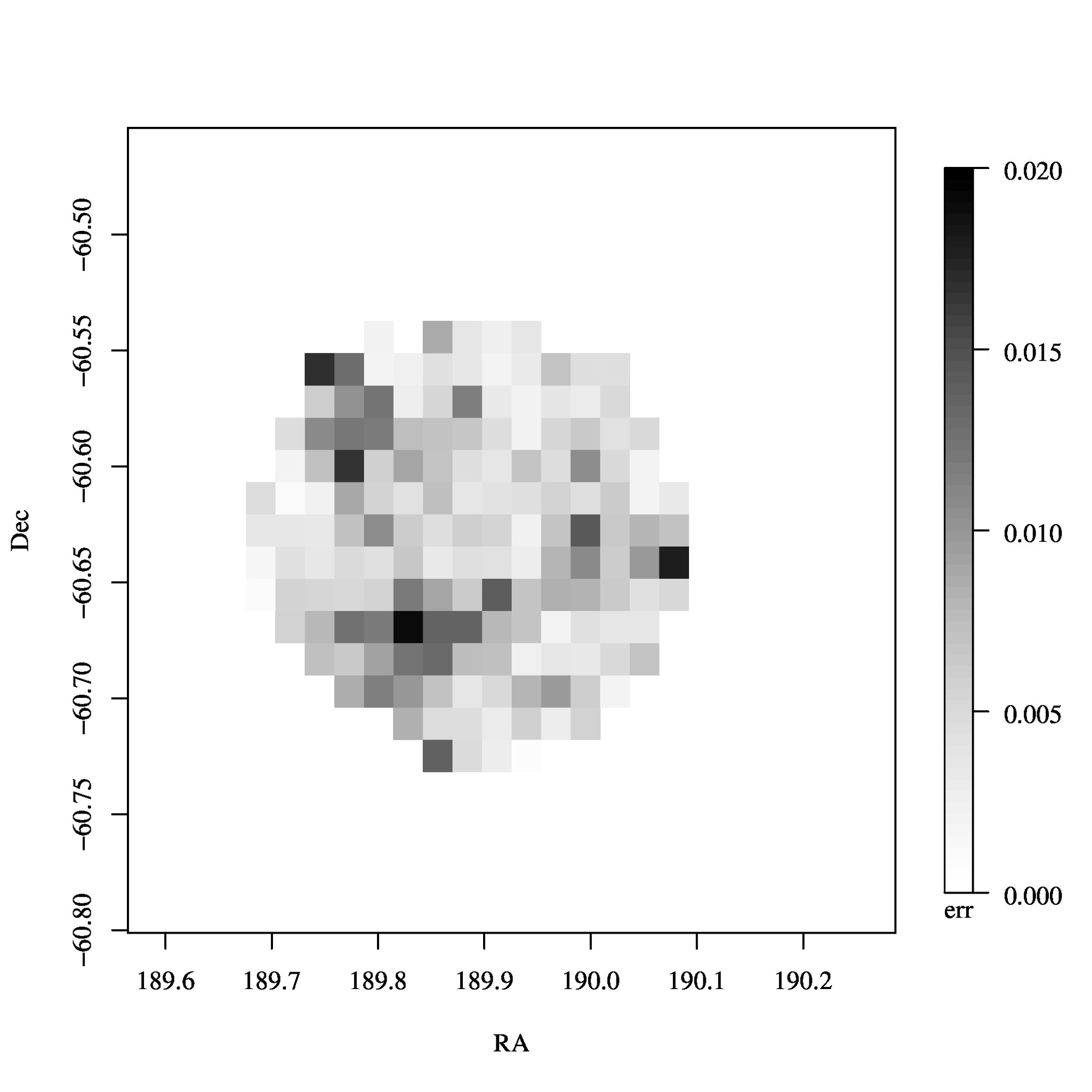} 
\caption{\label{fig:drgrid}\textit{Upper panel:} Colour deviations 
from the reference line due to the effect of DR, mapped on a $50\arcsec\times50\arcsec$ grid for 
stars inside 6$\arcmin$ from the centre. \textit{Lower panel:} Corresponding error map for the 
computed colour deviations from the reference line. The grayscale on the right side of each panel 
indicates the level of each parameter plotted.} \end{center}
\end{figure}

In Fig. \ref{fig:drgrid} (upper panel) we show the map of the DR obtained in terms of
 $\Delta E(B-V)$ with 
respect to the fiducial line. It ranges from about $-0.07$ to about +0.10. In particular, 
a region of low reddening is 
clearly identifiable. Comparing our results with those presented in Fig. 3 of P12, we obtain 
qualitatively the same result, with a region of lower DR in the north-western part. We found an 
excursion in the DR estimates of about 0.15 mag, similar to the 0.1 mag discussed in P12. Our higher 
value can be explained by the fact that we did not impose a blue envelope, allowing negative 
correction for DR, while P12  fixed the DR at zero for stars bluer than their reference line. 
In Fig. \ref{fig:drgrid} (lower panel) we show the corresponding map of the error associated to our estimates. The discrete 
appearance of these maps is due to two facts: the poorness in sampling a circular area using 
polygonal bins and the avoidance of interpolation in the corners, where the poor statistics may 
produce weak estimates. The table with the DR estimates is available through CDS.

\begin{figure}
\begin{center} \includegraphics[scale=0.45]{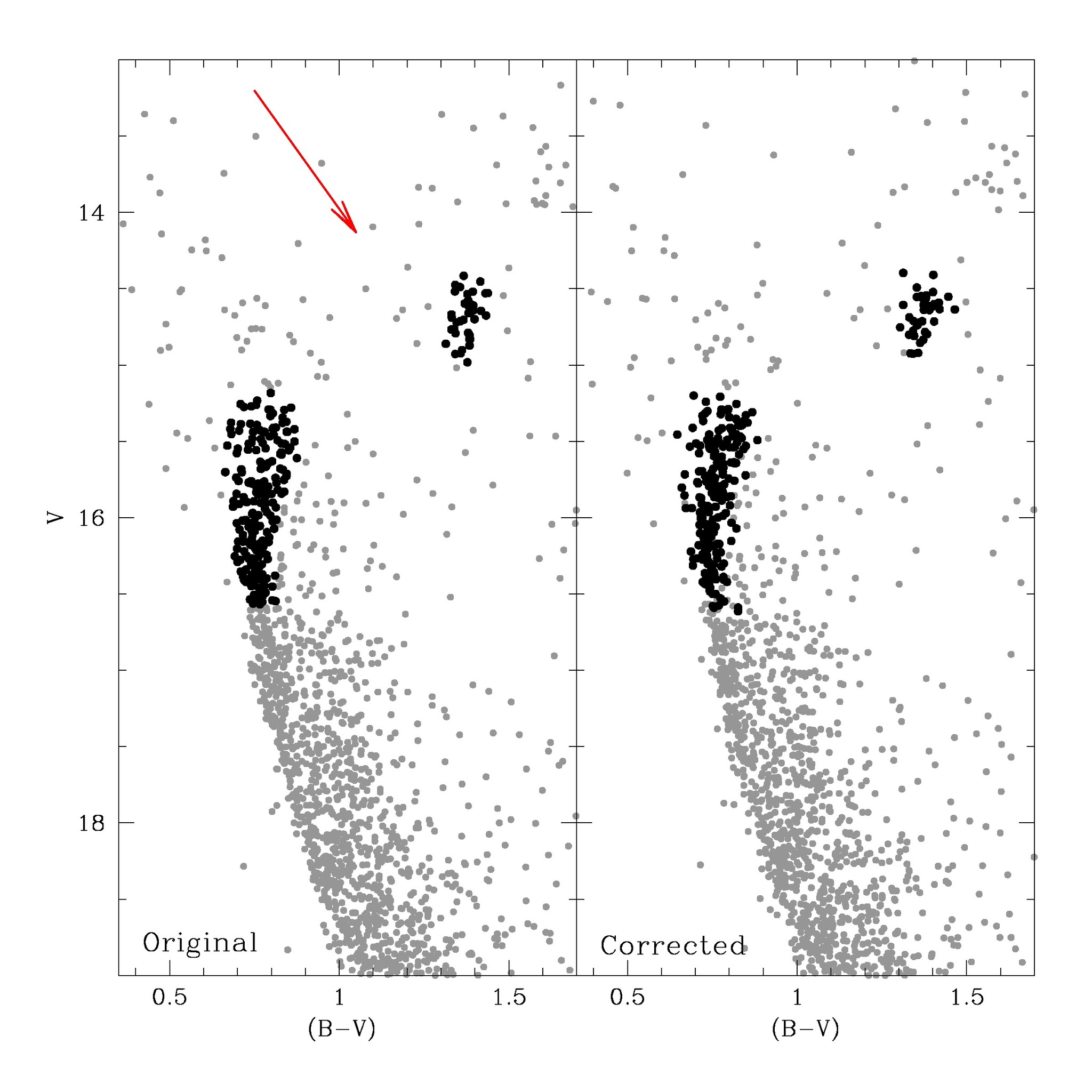} 
\caption{\label{fig:cmddr}CMDs for Tr~20 
stars (photometry from C10) with distance from the cluster centre $d<4\arcmin$. \textit{Left panel:} 
original photometry. The red arrow shows the direction of the reddening vector. \textit{Right panel:} 
photometry corrected for DR.
The TO and RC regions are plotted in black to clearly show the effect of the correction for DR.} \end{center}
\end{figure}

The overall effect of the DR on the CMD appearance is shown in Fig. \ref{fig:cmddr}. The MS and 
MSTO regions appear tighter, reducing the broadening of these phases substantially. This 
improvement is highlighted in black in the figure for the upper MS, but the lower part also benefits from 
the DR correction. The RC stars are more clumped, highlighting the peculiar morphology of this 
phase (see Sect. \ref{sec:rc}). Our DR estimate did not change the luminosity level and colour of age-sensitive indicators such as the MSTO, the bright limit of the MS, or the red-hook (RH) phase, limiting artificial 
estimates of the cluster parameters. The main difference to the method used in P12 is that we 
used many more stars to estimate the DR, which is therefore supported by a robust statistic. 
Moreover, we selected stars on the lower MS, avoiding objects at the MSTO. In this part of the CMD 
other physical mechanisms than DR have a more significant impact on the star magnitude and 
colour (in particular binaries), and, furthermore, the shape of the MS is much more sensitive to the 
metallicity and age, limiting the accuracy on both the DR estimates and the definition of the 
fiducial. On the other hand, the P12 method has the advantage of using only cluster members. Even 
though we have spectra for 1370 stars, we selected only 520 candidate members (see Sec. \ref{sec:rv}) and 100 fall in the MS 
box. They are still too few for a statistically significant estimate of the DR on the cluster field. However, 
we can quantify the differences between the two methods for the stars in common. Adopting the 
same MS box and fiducial, we compared the DR corrections obtained for single stars using our 
method and that of P12, without applying spatial smoothing. We found that the average difference 
is -0.004 mag with a dispersion of 0.02 mag. No systematic differences between the two methods were 
found, but only a low intrinsic dispersion. As final caveat, we stress that photometric errors, 
undetected binary systems, and residual contamination from the field might affect the DR estimation 
because they all produce a broadening of the MS. Our results are therefore an upper limit to the DR. 

\subsection[]{Red clump}\label{sec:rc}
\begin{figure}
\begin{center} 
\includegraphics[scale=0.45]{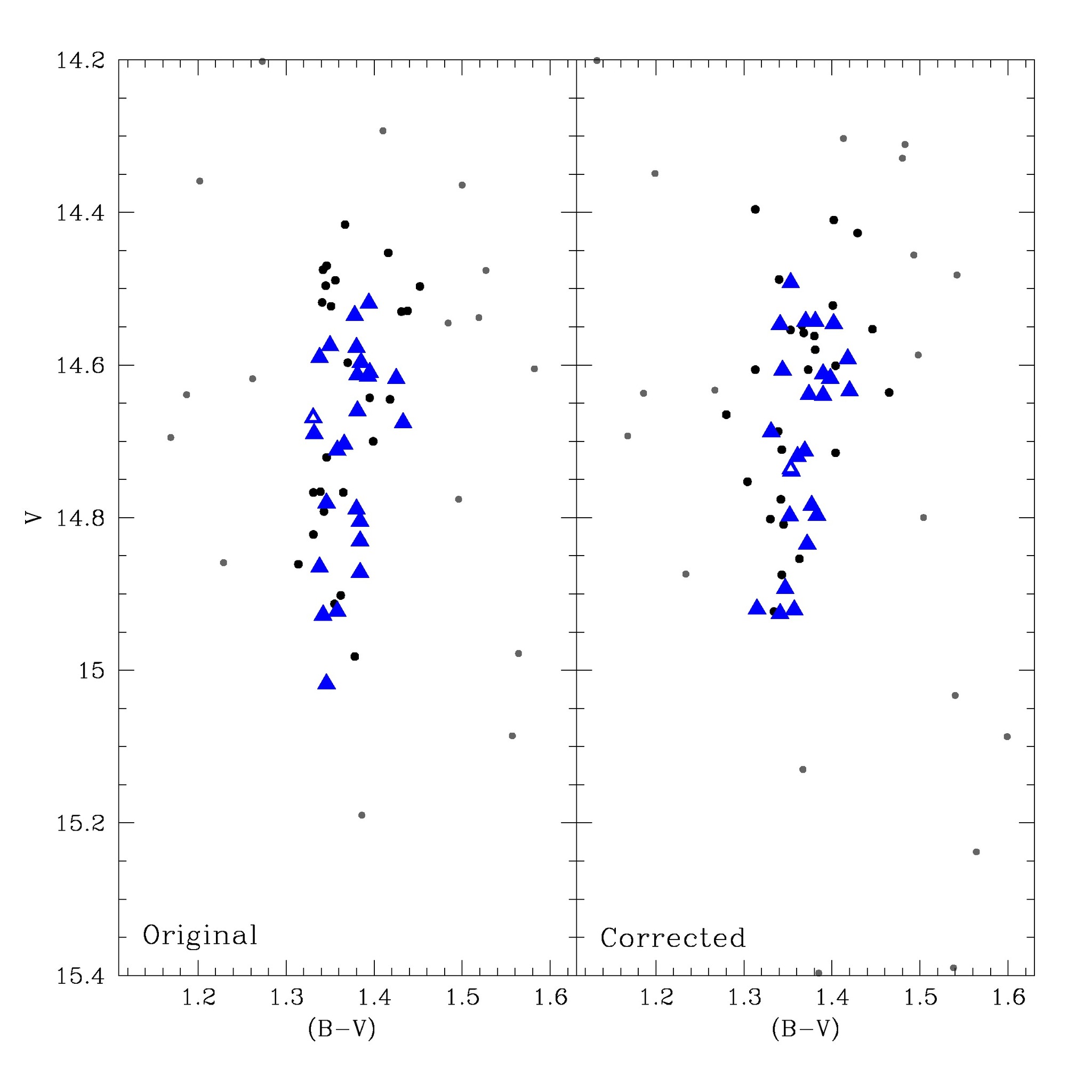} 
\includegraphics[scale=0.45]{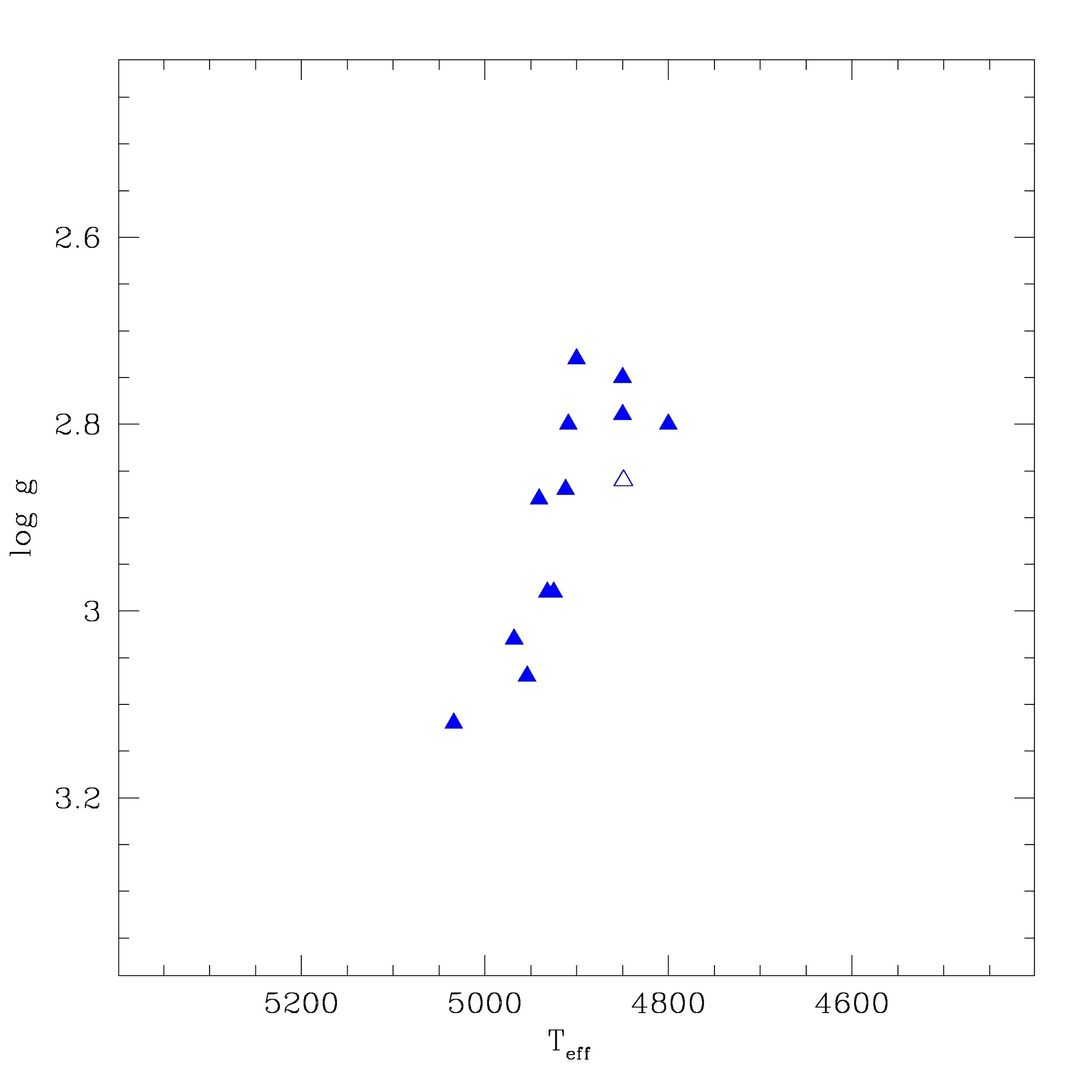}
\caption{\label{fig:rc}\textit{Upper panel:} As in Fig. \ref{fig:cmddr}, but for a CMD region centred on 
the RC position and for stars inside 5$\arcmin$. 
The blue triangles are the UVES candidate members. 
The open triangle is the outlier in chemical abundance (see Table \ref{tab:abuuves}).
\textit{Lower panel:} the UVES target candidate members in the theoretical  $\log g$ vs. $T_{eff}$ 
plane (only the 13 stars with complete analysis).}
 \end{center} 
\end{figure}

Tr~20 has been known to feature an extended RC at $B-V\simeq 1.45$, spread from $V\cong14.5$ 
to $V\cong15$. The de-reddened photometry of Figs. \ref{fig:cmddr} and \ref{fig:rc} shows that this 
extension is real and is not created by DR, as was found by P12. Furthermore, when DR is taken 
into account, the double structure of the RC, which has previously been discussed by C10, becomes 
more evident (see Fig. \ref{fig:rc}, upper panel). Two distinct groups of stars are evident, one extended and fainter, centred on about $V\sim14.8$, the other more luminous, centred on $V\sim14.5$.
We see, however, that the RC cannot be fitted by a single isochrone (see Sect. \ref{sec:param}). Table 
\ref{tab:abuuves} sums up the spectroscopic properties of the 13 UVES targets that are completely analysed;
we list the identification and $B$, $V$ magnitudes in C10, the Gaia-ESO Survey identifications, coordinates, RVs and the Gaia-ESO Survey atmospheric parameters $T_{eff}$, $\log g$, [Fe/H], and microturbulent velocity $\xi$ values.
 The numbers in Table \ref{tab:abuuves} confirm that the observed targets are giant stars, possibly in the RC phase, and 
are all very good candidate members for RV. Except for 
one star, which is within 3$\sigma$ from the average metallicity, they also show a remarkable 
chemical homogeneity, with an average iron abundance [Fe/H]$=+0.17$ and a dispersion of only 
0.03 dex. In Fig. \ref{fig:rc} (lower panel) we show the 13 UVES targets with abundance analysis in the theoretical plane $T_{eff},\log g$. They seem to have the same elongated shape as was found in the photometric plane.

Structured RC have been found in other OCs. 
The works of \cite{merm89,merm90} and \cite{merm97} present about ten intermediate-age MW OCs that
show this peculiarity. The first interpretation of these findings, confirmed by the subsequent works of 
\cite{gir99} and \cite{gir2000b}, is the possibility that some stars in the RC phase have undergone evolution 
through helium-core flash, while others have not, because of small differences in the exact core mass. 
Such differences require a considerable mass spread for clump stars of about 0.2 $M_{\odot}$, however, which could in principle result from different mechanisms:
\begin{itemize}
  \item the natural mass range of core-helium burning stars found in single isochrones, although the 
current models do not have the level of detail necessary to completely explore this possibility;
  \item a broad age spread (broader than 100 Myr), even if never observed in MW OCs;
  \item star-to-star variations in the mass-loss rates during the RGB phase. Recent asteroseismologic 
studies on the two OCs NGC~6791 and NGC~6819, the latter of which has an age and metallicity similar to Tr~20 
\citep[see][]{miglio12,corsaro12}, seem to indicate that no extreme mass-loss during RGB phases 
should be expected; 
  \item different stellar rotation history;
  \item dispersion in the overshooting efficiency in the convective core;
  \item binarity, if interaction and mass transfer or even mergers are considered.
\end{itemize}
Nevertheless, the transition between non-degenerate and degenerate He-core ignition would explain 
the findings in old OCs (age of about 1.4-1.6 Gyr), but not in other OCs, which are too young to be 
compatible with this evolutionary explanation. Tr~20 is an old cluster (see Sect. \ref{sec:param}) and 
agrees beautifully with this hypothesis. On the other hand, we cannot exclude the shape of the MSTO from this 
reasoning: even when we remove the effect of DR, a spread is still visible in the 
CMD that could be due to binary systems or to an age spread (more unlikely); 
both could give rise to a secondary RC. 

Alternatively, \cite{carraro11} studied NGC~5822 (0.9 Gyr old) and proposed that part of the stars of its 
apparently double RC might instead be RGB stars. With the abundance analysis of the UVES targets, 
the stars in the fainter part of the RC have a $\log g$ too large to be fit by the RC phase of the 
evolutionary models considered in this work. They might possibly be giant stars in the RGB phase, even if they are too 
warm for the best-fit model (see Sect. \ref{sec:param}, where we derive the cluster 
parameter with the isochrone fit). In this case, the accurate estimate of lithium abundance might be 
used to distinguish stars in the RGB phase 
(lithium should be lower for RC than for RGB stars), but at the moment we do not have this 
information for our targets.

Understanding the RC of Tr~20 is not easy at all. More investigations are needed to test 
all the possible hypotheses. About 30 additional giant stars have been observed with UVES, 
thus it will be possible to obtain a detailed abundance analysis and spectroscopic parameters 
for a significant fraction of stars that appear in the structured RC of the observational CMDs. Moreover, the Gaia-ESO Survey observations can be used 
to find binary systems (when combined with archive data), and examine 
the  effect of binaries on the observational CMD quantitatively.

Finally, we note that double RCs are also evident in clusters of the Magellanic Clouds as shown in \cite{glatt08} and \cite{mil09} and demonstrated in \cite{gir2009}. In these high-mass clusters the presence of stars of different age 
(due to a gap in 
star formation or a prolonged star formation) looks more probable, similarly to the
massive old globular clusters of the MW. However, the transition from 
non-degenerate and degenerate He ignition is not the universally accepted explanation for them, either. The effect of stellar rotation 
on the observational CMD (proposed e.g., by \citealt{bast09} or \citealt{li12}
and refuted by \citealt{gir11}) or of binary interactions  and merging \citep[e.g.][]{yang11} 
were considered to reproduce the double RCs. 

\section[]{Spectroscopic analysis}\label{sec:spectra} 
As we mentioned in Sec.~\ref{sec:spectro1}, the Gaia-ESO Survey data gathering and processing is
organised in WGs. In particular, stellar parameters and chemical abundances of 
F-G-K stars are derived by  WG 10 (for a description see Recio-Blanco et al., in prep.)
for GIRAFFE, and WG 11 derives them for UVES (Smiljanic et al., in prep.). 
Earlier-type stars, such as those found near the MSTO in Tr~20, are taken care of separately.
Within WG 10 and WG 11, the analysis  is performed by several 
nodes with different techniques. The common ground for each node is to assume 
local thermodynamic equilibrium  (LTE), using the same model atmospheres 
\citep[the MARCS models,][]{gust08}, the same grid of 
synthetic spectra (see \citealt{laverny12} and Recio-Blanco et al., in prep.), a line list with common atomic parameters 
(Bergemann et al., in prep.), and a common solar zero point \citep{grev07}. The results of the 
nodes are then combined to derive a final set of stellar parameters for each star. With this 
recommended set of stellar parameters, the nodes recompute the elemental abundances, which are 
combined to give a set of final values per star. More details can be found in the referenced papers.
However, we reiterate that we used only the RVs for the entire spectroscopic sample
and the abundances for 13 RC stars observed with UVES.

\subsection[]{RV distribution}\label{sec:rv}
\begin{figure}
\begin{center} 
\includegraphics[scale=0.45]{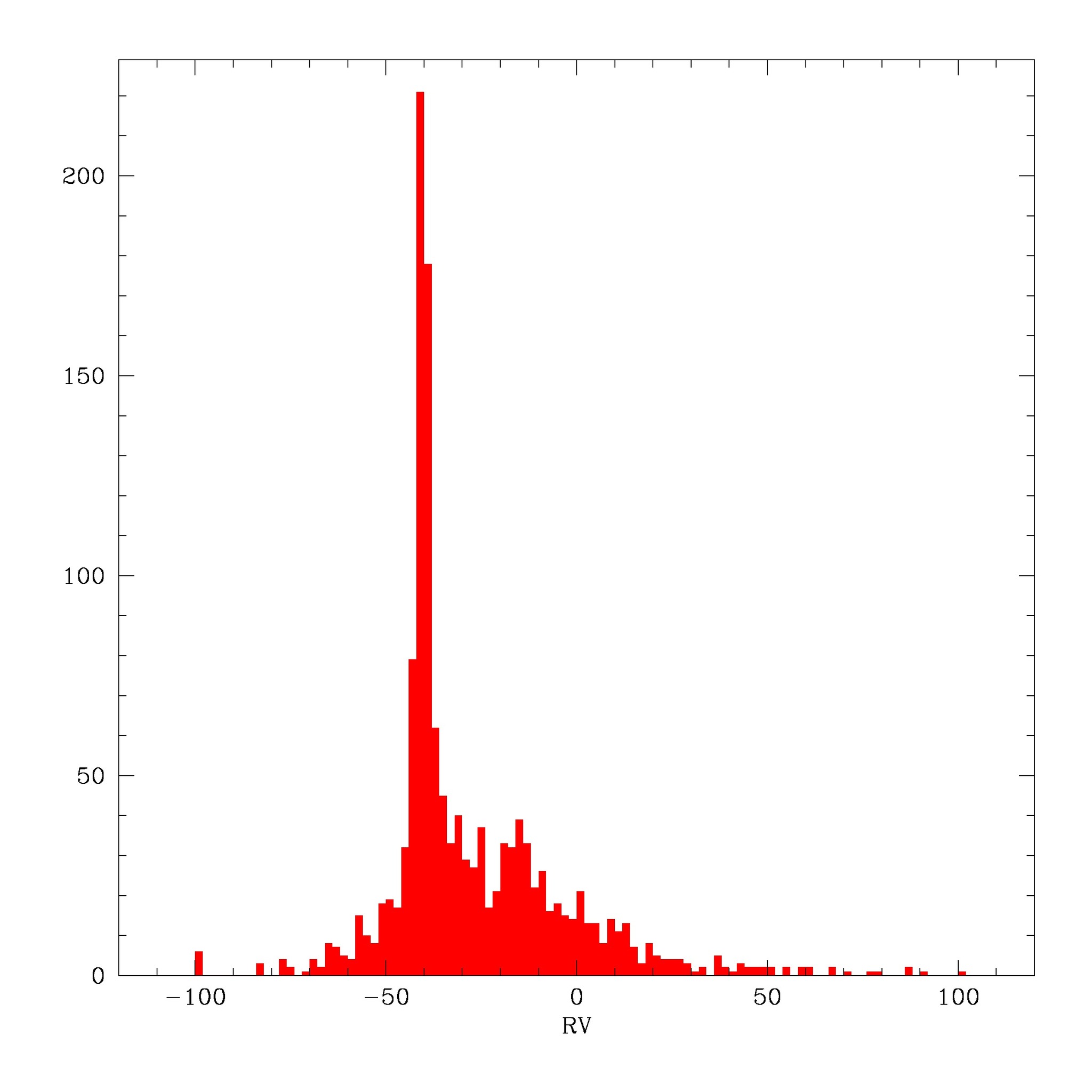} 
\caption{\label{fig:rvdista}RV distribution of all the GES 
targets (1370 stars). The peak of cluster stars is evident at about $-40$ km~s$^{-1}$.}
\end{center}
\end{figure}

Using all the RV measurements obtained for the stars observed in Tr~20, it is easy to identify the 
cluster signature with respect to the field stars. In Fig. \ref{fig:rvdista} we show the RV distribution for the entire Gaia-ESO Survey and archive targets (both UVES and GIRAFFE). The typical error on the RV for UVES 
targets is about 0.4 km~s$^{-1}$, while it ranges from 0.3  km~s$^{-1}$ to several 
km~s$^{-1}$ in the 
worst cases for GIRAFFE targets (see Table 3). 
We used stars in common between the setups to align the GIRAFFE RV estimates to that of 
UVES, finding that a systematic correction of about -0.46 km~s$^{-1}$ is needed for the HR15n 
spectra and of -0.50 km~s$^{-1}$ for the HR09b ones. Since we are not interested in 
the detailed cluster  internal dynamics, but only aim to identify candidate member stars, 
we did not try harder to homogenise the 
RVs, for instance, by using sky lines to correct for offsets between the zero points of individual spectra. Stars 
observed with different setups were considered using the following priorities: we used the UVES RV if 
available, the average RV between GIRAFFE setups when the star was observed with both HR15n and 
HR09b, and finally the RV derived from only one setup. We estimated the average RV of the sample by
selecting stars at different distances from the cluster centre to verify that consistent values were 
obtained. The inner part of the cluster has of course a higher percentage of cluster members than more distant fields, hence the estimate of the cluster average RV is more robust against spurious 
interlopers. On the other hand, statistics are poorer since the targets are spread up to 12$\arcmin$ 
from the centre. We decided to choose 8$\arcmin$ as the limiting distance for this analysis.

\begin{figure}
\begin{center} 
\includegraphics[scale=0.45]{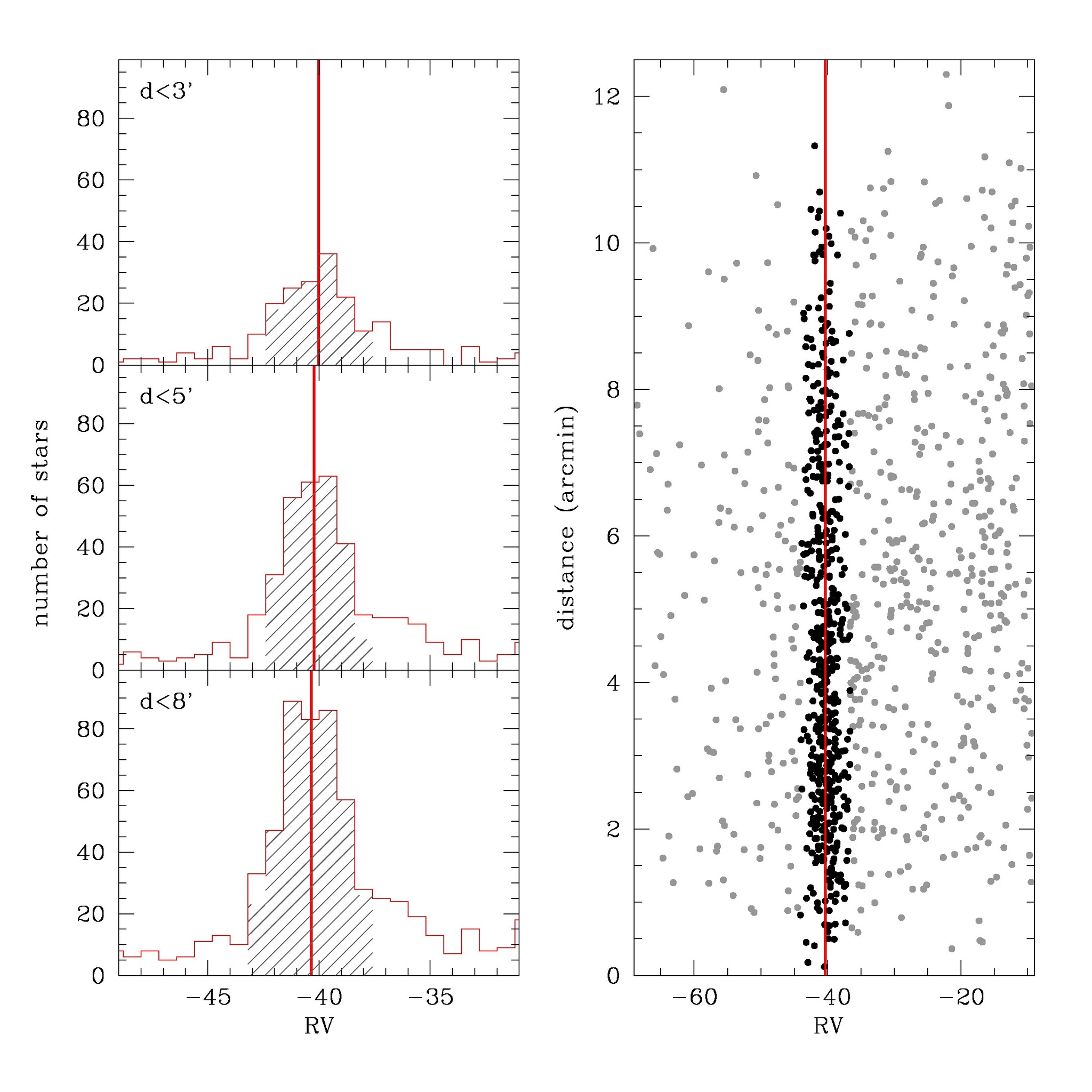} 
\caption{\label{fig:rvdist}\textit{Left panel:} RV distribution of 
all the GES targets for different distances from the cluster centre. The vertical line is the average RV 
of stars in the shaded histogram, obtained after an iterative two-sigma-clipping on the median. 
\textit{Right panel:} distance from the cluster centre versus RV of the targets. 
The black dots are the stars 
selected as candidate members using the average RV (red line) and $\sigma$ obtained from the 
whole group of targets (bottom histogram, left panel). }
\end{center}
\end{figure}

For each selection made on distances from the centre, outliers were expunged using the following 
method: the stars whose RV fell in the smallest RV interval containing 68\% (about the percentage of 
occurrences inside one standard deviation in a normal distribution) of the RV distribution were 
retained, then candidates were iteratively selected by using a two-sigma-clipping statistics on the median 
(five iterations were used). With the last selection convergence was reached and the average and the 
dispersion were computed. The results of this procedure are shown in Fig.~\ref{fig:rvdist} (left panel). We
stress here that our aim 
is to better define the evolutionary sequences on the CMDs by using only the most 
probable members. A 
more reliable membership estimation needs a more conservative approach that also considers 
binaries that might lie outside our selection criteria even if they belong to the cluster.  

The values obtained for different selections in distance agree within 0.2 km~s$^{-1}$; the estimate 
obtained in the inner region has a higher dispersion because of the small number of stars. 
The final values for the average RV and standard deviation used are those obtained for stars 
inside 8$\arcmin$. The average RV velocity is $\langle$RV$\rangle=-40.357\pm0.003$ km~s$^{-1}$ 
(the rms of the sample used to compute the average RV was $\sim$ 1.239 km~s$^{-1}$)\footnote{ 
The quoted dispersion is relative to the sample of stars selected to estimate the systemic velocity of 
the cluster (411 stars); it is indeed a lower limit of the expected dispersion of the cluster RV because in our 
analysis we did not consider the effect of binaries, for example.}. Considering only the 13 UVES targets 
with abundance analysis, we obtain an average $\langle$RV$\rangle=-40.26\pm0.11$ km~s$^{-1}$, in agreement with the whole sample of 
stars. For comparison, the average velocity found by P08 is $-40.8$ km~s$^{-1}$ (based on 
five stars),  while P12 obtained  $-40.40\pm0.12$ km~s$^{-1}$ by analysing 68 RC targets. 
Both agree very well with our values.

Our sample of targets encompasses MS and evolved stars, observed with two instruments of different resolution and spectral coverage.  
We defined the cluster candidate members using the following simple selection criterion: stars 
with an RV higher or lower than three times the rms with respect to the cluster average are not 
considered cluster 
members. One of the goals of the Gaia-ESO Survey is, in fact, to clean the sequences in the CMDs using 
membership information and combining Gaia-ESO Survey and archive spectra. From this we
found that $\sim$38\% of the targets (520 stars out of 1370) are good candidate members 
(see Fig.~\ref{fig:rvdist}, right panel). According to the Besan\c{c}on model \citep[see][]{robin03}, which was 
computed for the same coordinates of Tr~20, we estimated that indicatively\footnote{The
Besan\c{c}on model is not appropriate for small scales such as our case, but can still be used to 
obtain an approximate description of the field contamination.} about 17\% of the candidate 
members for RV may still be field stars. As shown 
in Figs. \ref{fig:gestgts} and \ref{fig:isoc}, almost all the targets located in the SGB region were 
discarded. The same has been found in P12 (see their Fig. 2), and confirms the expectations 
from theoretical 
models. The SGB phase has a very short timescale and is evidently very rich in clusters, but poorly 
populated in smaller clusters. We also found that many stars near the MSTO are not candidate 
members, as P12 did.

\subsection[]{Cluster metallicity}
\begin{figure}
\begin{center} \includegraphics[scale=0.45]{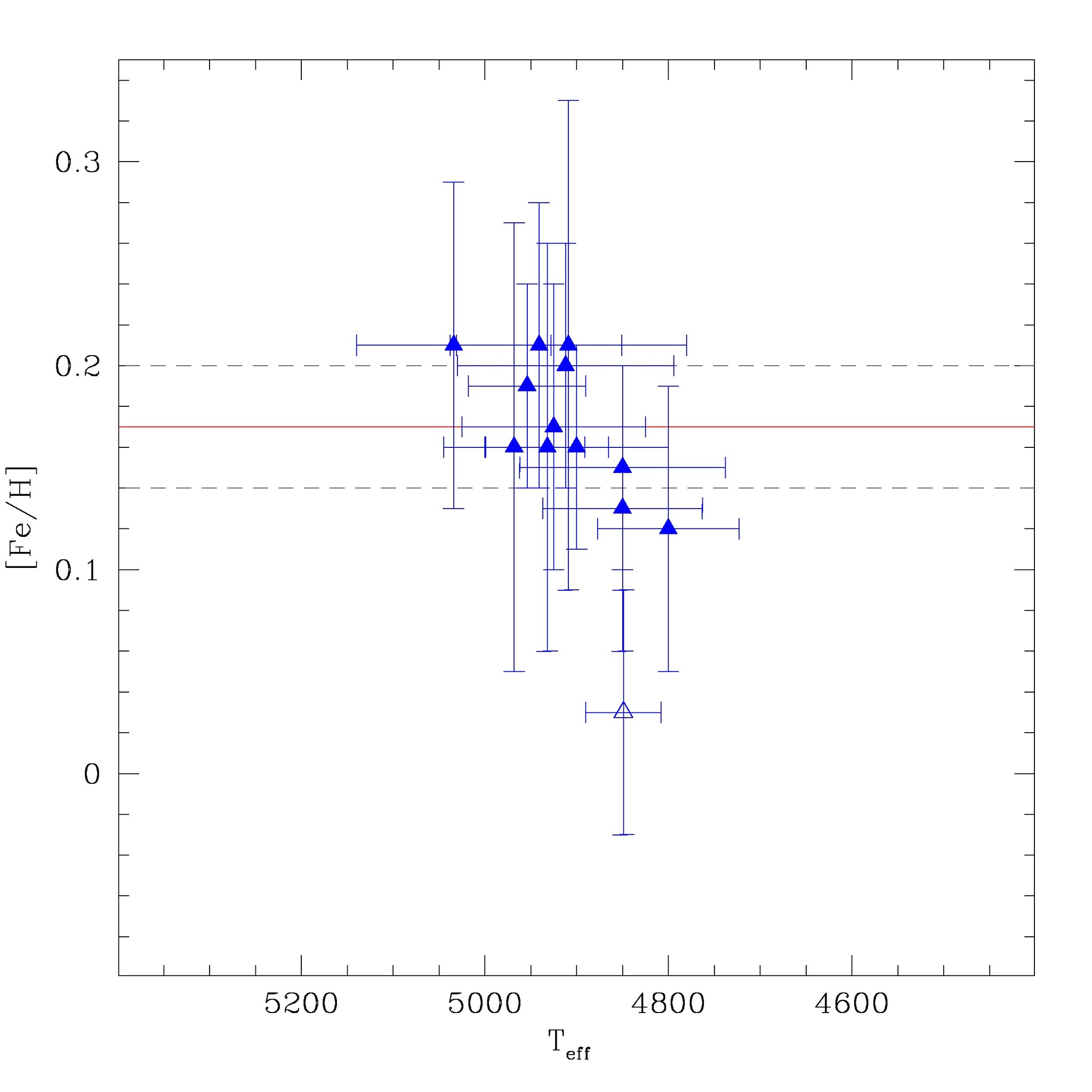} \caption{\label{fig:tfe}UVES targets in the T$_
{eff}$,[Fe/H] plane. The red line is the average metallicity; the dashed lines define the confidence 
interval.} 
\end{center}
\end{figure}

In Table \ref{tab:abuuves} we summarise the main results of the abundance analysis for the 
RC stars \citep[a detailed description is provided in][where  the metallicity and the abundance ratios of 
four $\alpha$-elements -Mg, Si, Ca, and Ti- and of three iron-peak elements -Fe, Ni, and Cr- are 
discussed]{mag13}. A metallicity [Fe/H] as recommended atmosphere parameter was considered here (see Smiljanic et al. in prep.). In Fig. \ref{fig:tfe} we show their [Fe/H] vs. $T_{eff}$; there seems to be a mild correlation, even though errors on quantities reduce 
the statistic significance. This minor effect is irrelevant for the goals of the present work, however, 
because we used the metallicity information only to confirm the membership and to choose the
appropriate isochrones (see next section). 
All 13 stars are very good candidate members according to their RV (see Sect. \ref{sec:rv}) and have a 
very low dispersion in metallicity. 
Only J12391577-6034406 (\#340) shows a slight discrepancy in the iron abundance relative to the other 
members. The average metallicity is $\langle$[Fe/H]$\rangle=0.17$ dex (with a dispersion of 0.03 dex) without this
star, and $\langle$[Fe/H]$\rangle=0.16$ (with a dispersion of 0.05 dex) with it. J12391577-6034406 (\#340) is within $3\sigma$ from the average. 
This places Tr~20 in 
the super solar metallicity regime. 

We recall that the solar abundances we adopted are those of 
\cite{grev07}, that is the iron abundance of the Sun is 7.45. Since this is the same value as 
adopted by P08, our metallicity is significantly higher than their estimate of $-0.11$ dex, which was based 
on only one RGB star. Unfortunately this star is not in common with the Gaia-ESO Survey UVES targets, 
which were all chosen to lie on the RC; this prevents a direct comparison and
investigation of systematics between the different analyses. On the other hand, the star analysed 
by P08 is  located in the upper part of the RGB, where 1D atmospheric models have more 
difficulties in reproducing real stars. In the past 
it has already been found that for low gravity and 
temperature the abundance analysis leads to lower 
iron abundances than for RC stars \citep[see, e.g.,][for a few examples]{friel03,carretta05}.

\begin{table*}
\setcounter{table}{3}
\small\addtolength{\tabcolsep}{-1pt}
\centering
\caption{ \label{tab:abuuves} Information for the 
13 UVES targets with complete abundance analysis. }
\begin{tabular}{rcccccccccc}
\hline
\hline
id & GES id & $V$ & $B$ & RA & Dec & RV & T$_{eff}$ & $\log g$ & [Fe/H] & $\xi$  \\
   &        & (mag) & (mag) & (deg) & (deg) & (km~s$^{-1}$) & (K) & (dex) & (dex) & (km~s$^{-1}$) \\
\hline
 340 & 12391577-6034406 &  14.67 &  16.00 & 189.8157333 & -60.5779569 &  -40.02$\pm$0.42 & 
4849$\pm$41 & 2.86$\pm$0.13 &  0.03$\pm$ 0.06 &  1.27$\pm$ 0.11 \\ 
 770 & 12392585-6038279 &  14.93 &  16.27 & 189.8577030 & -60.6410769 &  -42.54$\pm$0.42 & 
5034$\pm$106 & 3.12$\pm$0.31 &  0.21$\pm$ 0.08 &  1.25$\pm$ 0.10 \\ 
 505 & 12392700-6036053 &  14.52 &  15.91 & 189.8624847 & -60.6014733 &  -40.06$\pm$0.42 & 
4800$\pm$77 & 2.80$\pm$0.24 &  0.12$\pm$ 0.07 &  1.29$\pm$ 0.10 \\ 
 894 & 12393132-6039422 &  14.77 &  16.11 & 189.8804981 & -60.6617437 &  -35.98$\pm$0.42 & 
4954$\pm$64 & 3.07$\pm$0.15 &  0.19$\pm$ 0.05 &  1.20$\pm$ 0.22 \\ 
835 & 12393782-6039051 &  14.58 &  15.96 & 189.9075746 & -60.6514254 &  -40.53$\pm$0.42 & 
4909$\pm$129 & 2.80$\pm$0.21 &  0.21$\pm$ 0.12 &  1.30$\pm$ 0.16 \\ 
 346 & 12394419-6034412 &  14.70 &  16.07 & 189.9341452 & -60.5780773 &  -41.02$\pm$0.42 & 
4941$\pm$90 & 2.88$\pm$0.23 &  0.21$\pm$ 0.07 &  1.25$\pm$ 0.06 \\ 
 781 & 12394475-6038339 &  14.61 &  16.01 & 189.9364207 & -60.6427569 &  -39.28$\pm$0.42 & 
4850$\pm$112 & 2.75$\pm$0.22 &  0.15$\pm$ 0.05 &  1.38$\pm$ 0.08 \\ 
 791 & 12394596-6038389 &  14.54 &  15.91 & 189.9414399 & -60.6441467 &  -39.69$\pm$0.42 & 
4912$\pm$118 & 2.87$\pm$0.21 &  0.20$\pm$ 0.06 &  1.27$\pm$ 0.15 \\ 
 287 & 12394690-6033540 &  14.78 &  16.13 & 189.9453843 & -60.5650194 &  -40.98$\pm$0.42 & 
4968$\pm$77 & 3.03$\pm$0.10 &  0.16$\pm$ 0.11 &  1.14$\pm$ 0.14 \\ 
 795 & 12394742-6038411 &  14.71 &  16.07 & 189.9475589 & -60.6447566 &  -39.82$\pm$0.42 & 
4900$\pm$100 & 2.73$\pm$0.23 &  0.16$\pm$ 0.05 &  1.21$\pm$ 0.15 \\ 
 787 & 12395426-6038369 &  14.60 &  15.98 & 189.9760346 & -60.6436069 &  -42.45$\pm$0.42 & 
4925$\pm$100 & 2.98$\pm$0.13 &  0.17$\pm$ 0.07 &  1.36$\pm$ 0.04 \\ 
 399 & 12395975-6035072 &  14.62 &  16.04 & 189.9988948 & -60.5853363 &  -42.01$\pm$0.42 & 
4850$\pm$87 & 2.79$\pm$0.19 &  0.13$\pm$ 0.07 &  1.29$\pm$ 0.11 \\ 
1044 & 12400278-6041192 &  14.99 &  16.37 & 190.0115263 & -60.6886598 &  -39.04$\pm$0.42 & 
4932$\pm$67 & 2.98$\pm$0.11 &  0.16$\pm$ 0.10 &  1.37$\pm$ 0.05 \\ 
\hline
\end{tabular}
\end{table*}

\section[]{Cluster parameters}\label{sec:param}
By means of the considerable improvements obtained i) in the photometry with the DR estimation, ii) in 
the metallicity measurement with the high-resolution spectroscopy, and iii) in the 
membership with RV 
determinations, we derived the age, distance, and average reddening of Tr~20 by using the classical 
approach with isochrone fitting. We adopted three different sets of isochrones to have a less 
model-dependent solution for the cluster parameters: the PARSEC \citep{parsec},
 the BASTI \citep{basti}, and the Victoria-Regina \citep{victoria} isochrones.  

\begin{figure*}
\begin{center} \includegraphics[scale=0.90]{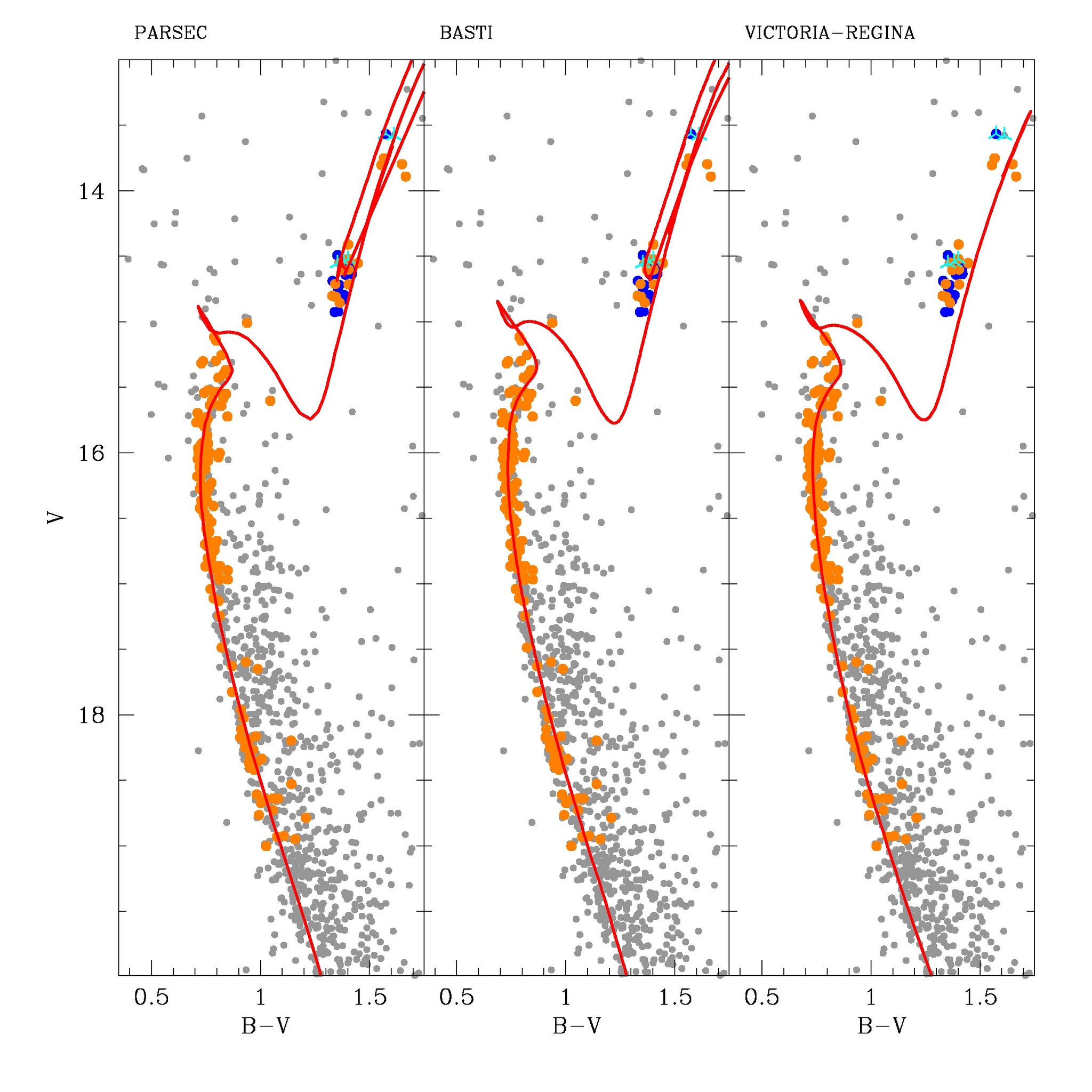} \caption{\label{fig:isoc}CMD obtained for 
stars inside 3$\arcmin$ using the photometry from C10 corrected for DR, and the best isochrone fit for different evolutionary models (PARSEC - left; BASTI - middle; Victoria-Regina - right). GES target non-members 
have been disregarded, while members are highlighted with orange (GIRAFFE) and blue (UVES) 
points. Cyan points are P08 members. See Table \ref{tab:isoc} for the adopted parameters for the 
isochrone fitting.} 
\end{center}
\end{figure*}

The best-fitting isochrone was chosen by eye examination as that which can describe the main 
age-sensitive evolutionary phases at the same time: the luminosity and colour of the MSTO, RH, and 
RC when possible. 
We used the metallicity resulting from the spectroscopic estimate, that is [Fe/H]=+0.17, and we 
converted it to Z taking into account the different solar abundances of the three sets of isochrones. 
The errors on the estimated parameters are mainly due to the uncertainties in the definition of the 
age indicators. In particular, the RC has a very peculiar morphology (see Sect. \ref{sec:rc}), which 
drives the main uncertainty on the age. The photometric error does not have a significant impact on 
the error budget, apart from systematic ones that can stem from the photometric data reduction and 
calibration. For instance, the differences found in magnitude and colour between P08 and C10 have 
an impact on the determination of distance modulus and reddening that will be discussed. For 
homogeneity we derive the cluster parameters using the catalogue from C10 corrected for DR in the 
$B-V$ colour and discuss the effect of the offset between the photometric data. In Fig. \ref{fig:isoc} 
we show the results of the best fits, summarised in Table \ref{tab:isoc}.

\begin{table*}
\centering
\caption{ \label{tab:isoc} Results, errors, and estimated systematic uncertainties using different evolutionary models with [Fe/H]$\simeq$+0.17. }
\begin{tabular}{lccccccc}
\hline
\hline
 Model & age & $(m-M)_0$ & $E(B-V)$ & $d_{\odot}$ & $R_{GC}$ & z & $M_{TO}$ \\
       & (Gyr) & (mag) & (mag) & (kpc) & (kpc) & (pc) & ($M_{\odot}$)\\
\hline
PARSEC & 1.66$\pm$0.2 & 12.64$\pm$0.1(0.2) & 0.32$\pm$0.02(0.05) & 3.37(0.3) & 6.87(0.02) & 130.71(10) & 1.8\\
BASTI & 1.35$\pm$0.2 & 12.72$\pm$0.1(0.2) & 0.31$\pm$0.02(0.05) & 3.50(0.3) & 6.86(0.02) & 135.62(10) & 1.9\\
VICTORIA & 1.46$\pm$0.2 & 12.70$\pm$0.1(0.2) & 0.35$\pm$0.02(0.05) & 3.47(0.3) & 6.86(0.02) & 134.37(10) & 1.9\\
\hline
\end{tabular}
\end{table*}

For the PARSEC set (see Fig. \ref{fig:isoc}, left panel), for which Z$_{\odot}=0.015$, we adopted Z=0.022. We found that the best age 
estimate is $1.66\pm0.2$ Gyr, with a reddening $E(B-V)=0.32\pm0.02$ mag, and a true 
distance modulus $(m-M)_0=12.64\pm0.1$ mag. The luminosity and colour of the RC are well 
reproduced, as are those of the RH and upper MS. As has often been found, the lower part of the MS, for 
$V>18$, is slightly redder in the model than in the observations.

For the BASTI isochrones (see Fig. \ref{fig:isoc}, middle panel) there is a coarse grid in chemical composition, hence we chose 
two different metallicities,  Z=0.019 (i.e. solar) and Z=0.03, which bracket our spectroscopic metallicity. 
The best-fit of RH and RC is obtained for an age of $1.35\pm0.2$ Gyr. The more metal-rich solution 
reproduces the RGB and RC phases slightly better, hence the adopted parameters are from this 
isochrone: $E(B-V)=0.31\pm0.02$ and $(m-M)_0=12.74\pm0.1$. 

The Victoria-Regina isochrones do not include the evolved phases after the RGB for the age 
of Tr~20. Therefore we chose 
the best-fitting solution as the one that best matches the MS and MSTO morphologies (see Fig. \ref{fig:isoc}, right panel). 
Moreover, it is 
only possible to use a coarse grid in terms of metallicity, hence we tried two different metallicity 
values that bracket our estimate: [Fe/H]=0.13 and [Fe/H]=0.22. They both fairly well 
reproduce the 
MS ridge-line for an age of $1.46\pm0.2$ Gyr and describe the bending at the MSTO and the 
slope of the lower MS very well, although we prefer that of the lower metallicity because its RGB is closer to the 
observations, while the latter has a redder RGB phase. 
The average reddening and distance modulus 
for this isochrone are $E(B-V)=0.35\pm0.02$ mag, and $(m-M)_0=12.72\pm0.1$ mag. With respect to 
other sets we found that this one reproduces the MS better, though the colour of the RGB is 
redder than observed. Furthermore, the age is more loosely constrained, since we lack the RC 
phase. 

We repeated this analysis with the Dartmouth isochrones \citep{dott08} and reached the same conclusions,
but for brevity we decided not to detail the analyses as for the other evolutionary tracks.
In summary, we found a nice agreement in average reddening and distance modulus for all 
the model sets. They are similar to what was found by C10. We set the age in 
the range from 1.35 Gyr to 1.66 Gyr. 

Using instead the P08 data, we expect to find some differences in distance modulus and reddening,
because the age is mainly constrained by the magnitude difference between the TO and RC 
luminosities. 
We found that $E(B-V)$ is about 0.05 mag lower and the distance modulus consequently 0.16 mag 
higher, which translates into a larger heliocentric distance (by about 0.3 kpc) and a greater height 
above the MW disc (about 10 pc). The differences with the P08 estimates arise because they 
used stellar models with subsolar metallicity (see Table \ref{tab:lit}); in particular, this explains their 
higher reddening. 

The results obtained with the three sets are given in Table~\ref{tab:isoc}, where we indicate
age, distance modulus, reddening, distance from the Sun and the Galactic centre (the 
$R_{GC,\odot}$  adopted is 8 kpc \citep[see][]{mal13}), distance from the Galactic plane, 
and mass of the stars at the MSTO. In parenthesis we quote the systematic errors as an additional uncertainty due to the zero points between the photometric catalogues of C10 and P08.

\begin{figure}
\begin{center} \includegraphics[scale=0.45]{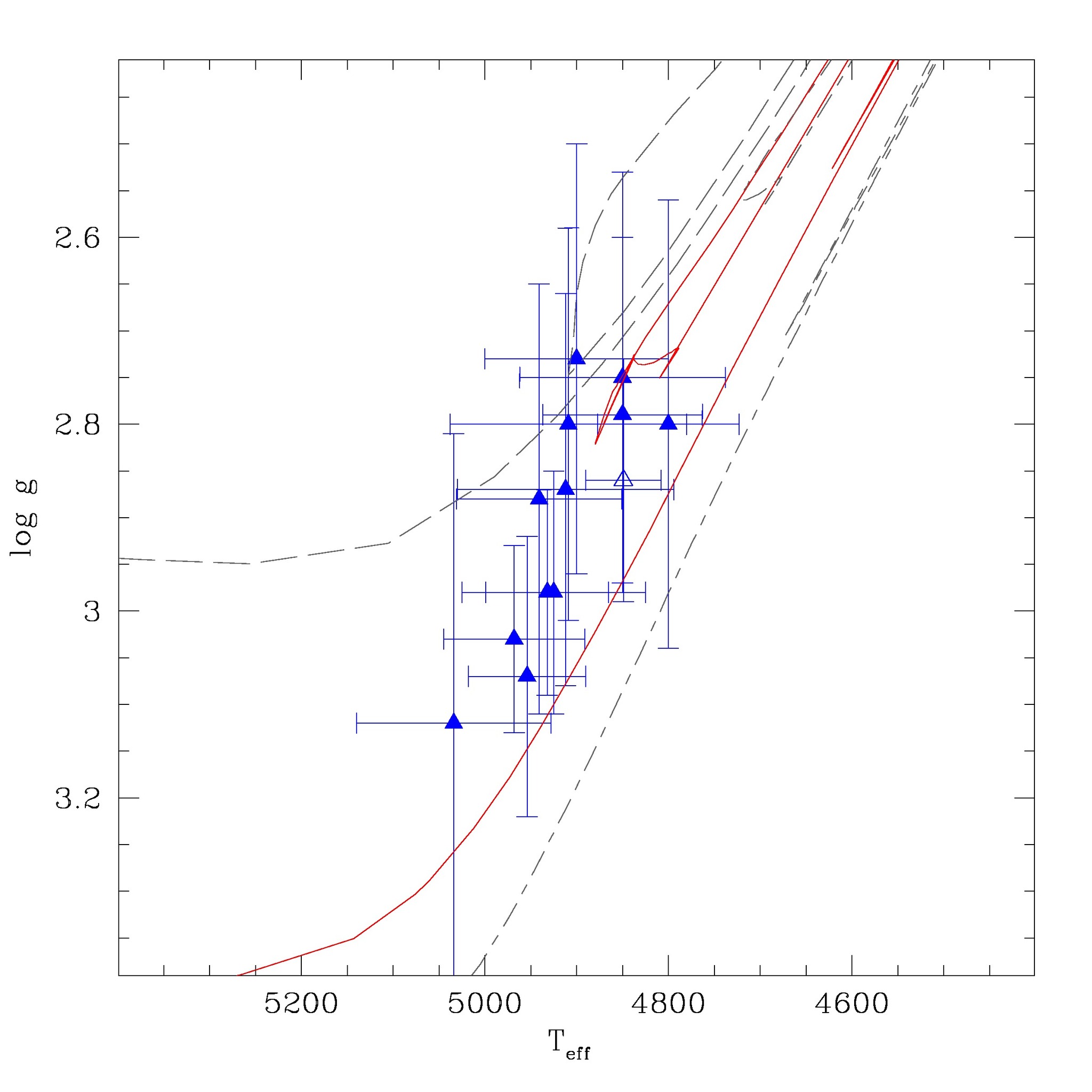} \caption{\label{fig:hr}Blue triangles are 
UVES targets. The red line is the PARSEC isochrone for the age 1.66 Gyr. For younger ages (dashed 
line, age of 0.6 Gyr) or older ones (dot-dashed line, age of 2.5 Gyr) the RC phase moves toward 
lower $\log g$ values with respect to the data.} 
\end{center}
\end{figure}

With the 13 UVES targets it is possible to evaluate the agreement between the models and the data 
in the theoretical plane $T_{eff},\log g$ for the first time for this cluster. In Fig. \ref{fig:hr} 
we show the best-fitting isochrone for the PARSEC set (continuous line): the RC phase 
of the model fits 
the upper clump of stars quite well, while the lower elongated group cannot be fitted by any age at this metallicity. 
In the same figure we show for comparison a younger and an older age isochrone 
(dashed and dot-dashed lines, respectively), for which the clump phase is located at lower 
$\log g$ with respect to the 
data, and never reaches $\log g>3$. These stars might be RGB stars (even if still too warm for the 
best-fitting model) instead of RC stars because they appear in the observational CMD. 
We discuss the peculiar structure of the RC of Tr~20 in detail in Sect. \ref{sec:rc}.

\subsection{Fitting $B-V$ and $V-I$}
\begin{figure}
\begin{center} \includegraphics[scale=0.45]{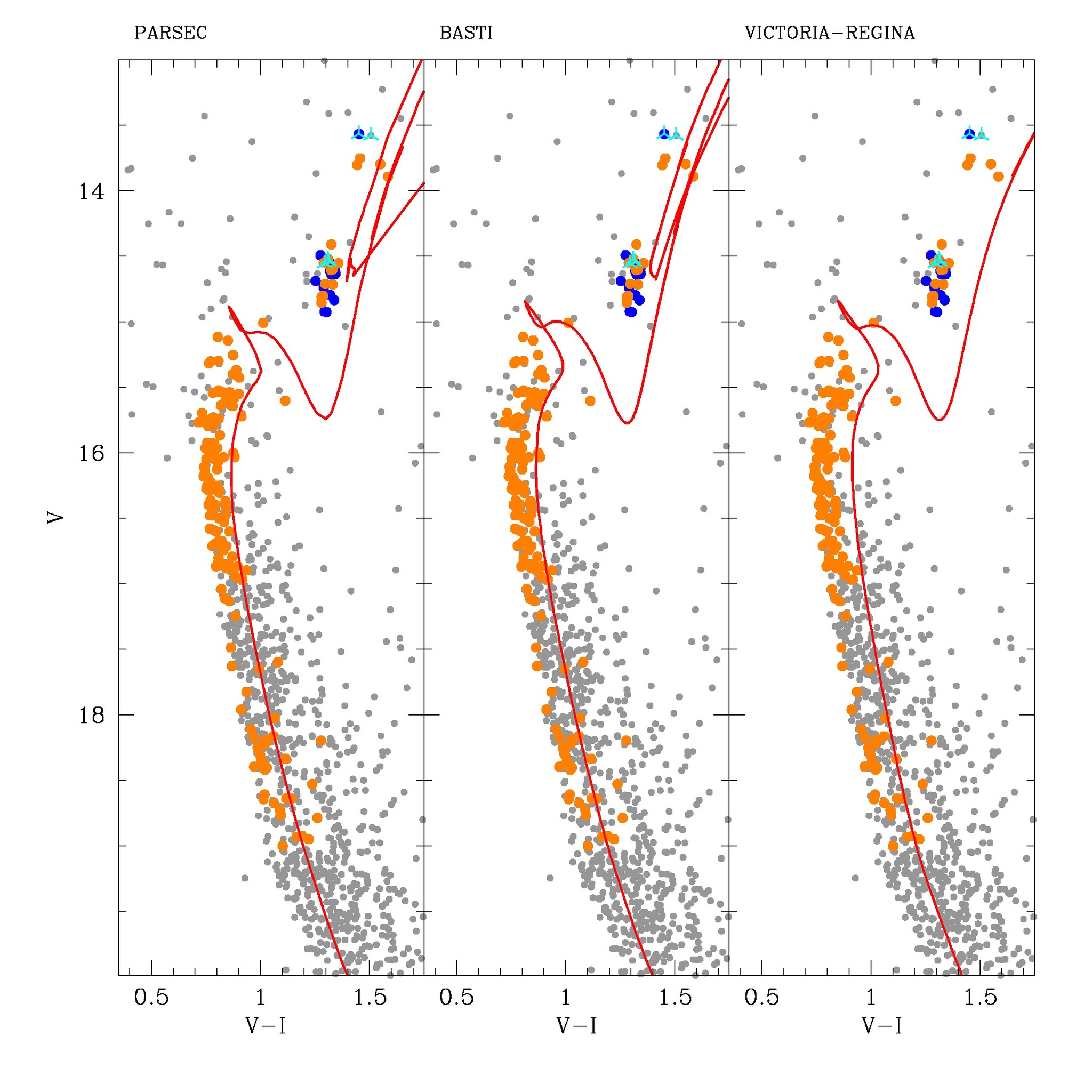} \caption{\label{fig:isoc_vi} Same as in Fig. 
\ref{fig:isoc}, but in the $V$ and $V-I$ CMD, after the standard extinction law has been applied for the 
isochrones.} 
\end{center}
\end{figure}

Multi-band photometry can be used to estimate the expected cluster metallicity; in principle,
the correct metallicity is the one that produces a good fit 
with the same isochrone in two different colours with the same parameters \citep[e.g.][]{tosi07}. It is 
very interesting to compare the metallicity obtained from photometry with the one estimated from 
spectroscopy. Unfortunately, by applying the standard extinction law to convert $E(B-V)$ to $E(V-I)$ ($E(V-I)
=1.25\times E(B-V)$),
we obtain a poor fit in $V-I$  for the spectroscopic metallicity, after the cluster parameters are 
fixed in $B-V$. 
In Fig. \ref{fig:isoc_vi} we show the ``best-fitting'' isochrone in the $V$, $V-I$ plane 
(C10 photometry) after applying 
the standard extinction law. The same inconsistencies hold, but in the opposite way, for the P08 
photometry. We tried to find the photometric metallicity that allows a match in both colours. For C10 
we derived with a very high metallicity, Z=0.05 or [Fe/H]$\sim$0.5, which seems implausible even 
for a cluster in the inner disc such as Tr~20. 
For the photometry of P08 a match was obtained with 
subsolar metallicity, Z=0.01 or [Fe/H]$\sim$-0.18 dex. Both metallicities are in contrast with the 
accurate spectroscopic value discussed in this paper.

On the other hand, since Tr~20 is located in the disc between spiral arms, it might be that the standard extinction law is no longer a good approximation. Using a different relation, such as $E(V-I)=1.62\times E(B-V)$ \citep[see][]{card89}, the agreement in the case of C10 photometry worsens, 
while for P08 we obtain a good match for the PARSEC and BASTI isochrones, but not for the Victoria 
isochrones. 

This failure in simultaneously fitting the CMDs in two colours has been found in other 
cases \citep[see e.g.,][]{ahu13}, but no definitive conclusion has been reached. Moreover, in our case 
we cannot firmly 
explain the poor match of stellar models in the $B-V$ and $V-I$. One answer can be the 
already known problem of the photometric transformations from the theoretical to the observational 
plane. The three models use different transformations, which adds a source of uncertainties to these 
comparisons. We cannot exclude problems related to the calibration of the 
photometric data, however, because the catalogues show systematic differences (Fig. \ref{fig:compareV}). 
Lacking better choices, we decided to constrain our analysis by only using the $B-V$ photometry 
(for which C10 and P08 agree better) and the metallicity from the Gaia-ESO Survey.

\section[]{Summary and conclusions}
We used available photometry from the literature and spectroscopic Gaia-ESO Survey data to make a 
comprehensive study of Tr~20 as a pilot analysis for all the old OCs in the Gaia-ESO Survey. We derived the 
cluster structural parameters and estimated the effect of DR for the inner part of the cluster. We 
found that extinction can vary significantly across the field of the cluster, with a range of more than 
0.1 mag in $E(B-V)$. The accurate abundance analysis of the Gaia-ESO Survey high-resolution spectra of 13 stars in the RC 
phase gives an average cluster metallicity of [Fe/H]=+0.17. From the RV distribution of 1370 stars we 
estimated the average radial velocity of the cluster, which is  $\langle RV \rangle=-40.36$ km~s$^
{-1}$, and defined the candidate member stars on the basis of their RV. With this information we 
were able to partially clean the catalogue of obvious non-member stars.

With this information and using the C10 photometry we estimated the age, distance, and reddening of 
the cluster by means of the classical isochrone-fitting approach. Using different models (PARSEC, 
BASTI, and Victoria-Regina), we found a cluster age in the range of 1.35-1.66 Gyr, an average 
reddening $E(B-V)$ in the range 0.31-0.35 mag, and a distance modulus $(m-M)_0$ in the range 
12.64-12.72 mag. Had we used the P08 photometry, these values would be 0.05 mag lower 
in $E(B-V)$ and 0.16 mag higher in $(m-M)_0$. This demonstrates the influence of using 
different photometric data sets, 
even when they are apparently of good quality. Only homogeneity can provide the best (internally consistent) 
parameters and avoid systematics. We cannot fit both $B-V$ and $V-I$ with the same model (with neither photometric set) for metallicities that reasonably agree with the spectroscopic one, a 
problem possibly related to the photometric transformations adopted for the theoretical isochrones or 
to systematic errors in the photometry of the two data sets. 

Our parameters agree reasonably well with most literature values, but were derived through a 
more robust method. We have discussed the problems of the RC of Tr~20, which cannot be fitted by a single isochrone, but 
have found no firm conclusion. More solid results will be obtained with the next data release, where all the Gaia-ESO Survey spectra of this cluster (especially the high-resolution UVES spectra) will be analysed producing a better measurement of the chemical abundances and metallicity and a deeper insight into the properties of the stars that appear to be in the RC phase.
This feature is interesting and deserves more investigation.

\section*{Acknowledgements}
This research has made use of the WEBDA database, originally developed by J-C. Mermilliod, now 
operated at the Department of Theoretical Physics and Astrophysics of the Masaryk University, and of 
BaSTI web tools.

Extensive use has been made of NASA's Astrophysics Data System Bibliographic Services, and of 
the SIMBAD database and VizieR catalogue access tool, CDS, Strasbourg, France. 

The results presented here benefitted from discussions in three Gaia-ESO workshops supported by 
the European Science Foundation through the Gaia Research for European 
Astronomy Training Research Network Program (Science meetings 3855, 4127 and 4415). 
We acknowledge the support from INAF and Ministero dell'Istruzione, dell'Universit\`a e della Ricerca 
(MIUR) in the form of the grant ``Premiale VLT 2012''. 
This work was partially supported by the Gaia Research for European Astronomy Training (GREAT-
ITN) Marie Curie network, funded through the European Union Seventh Framework Programme 
[FP7/2007-2013] under grant agreement no 264895.

D.G. gratefully acknowledges support from the Chilean BASAL Centro de Excelencia en Astrofisica y Tecnologias Afines (CATA) grant PFB-06/2007.

I.S.R. gratefully acknowledges the support provided by the Gemini-CONICYT project 32110029.

This work was partly supported by the European Union FP7 programme
through ERC grant number 320360 and by the Leverhulme Trust through grant
RPG-2012-541.

We thank Paolo Montegriffo (INAF, Osservatorio Astronomico di Bologna, Italy) for his software 
CataPack.

\end{document}